\documentclass[a4paper,onesided,12pt]{report}
\usepackage{styles/fbe_tez}
\usepackage[utf8x]{inputenc} 

\usepackage{amsmath, amsthm, amssymb}

\usepackage{tablefootnote}
\usepackage{subfigure}
\usepackage{tabularx}
\usepackage{adjustbox}
\usepackage{array}
\usepackage{pgfplots}
\usepackage{times}
\usepackage{multirow}
\usepackage{cancel}
\usepackage{tikz}
\usetikzlibrary{positioning,calc,decorations.pathreplacing}
\pgfplotsset{compat=1.5}
\usepackage{epic,eepic}

\usepackage{mathtools}
\DeclarePairedDelimiter\ceil{\lceil}{\rceil}
\DeclarePairedDelimiter\floor{\lfloor}{\rfloor}

\usepackage{algorithm}
\usepackage{algpseudocode}
\usepackage{caption}

\usepackage[symbol]{footmisc}
\usepackage{cite}
\usepackage{graphicx}
\usepackage{longtable}
\graphicspath{{figures/}} 

\usepackage{multirow}
\usepackage{subfigure}
\usepackage{booktabs}

\title{DESIGN, IMPLEMENTATION, AND ANALYSIS OF\\FAIR FAUCETS FOR BLOCKCHAIN ECOSYSTEMS}
\turkcebaslik{BLOKZİNCİRİ EKOSİSTEMLERİ İÇİN ADİL MUSLUK\\TASARIM, UYGULAMA VE ANALİZİ}
\degree{B.A., Psychology, Boğaziçi University, 2007\\
	M.A., Cognitive Science, Boğaziçi University, 2013}
\author{Serdar Metin}
\program{Computer Engineering}
\subyear{2022}

\supervisor{}
\examineri{}
\examinerii{}
\examineriii{}
\examineriv{}
\dateofapproval{DD.MM.YYYY}

\begin{document}

\pagenumbering{roman}
\makephdtitle 

\begin{abstract}
The present dissertation addresses the problem of fairly distributing shared resources in non-commercial blockchain networks. Blockchains are distributed systems that order and timestamp \textit{records} of a given network of users, in a public, cryptographically secure, and consensual way. The records, which may in kind be events, transaction orders, sets of rules for structured transactions etc. are placed within well-defined datastructures called blocks, and they are linked to each other by the virtue of cryptographic pointers, in a total ordering which represents their temporal relations of succession. The ability to operate on the blockchain, and/or to contribute a record to the content of a block are shared resources of the blockchain systems. In commercial networks, these resources are exchanged in return for fiat money, and consequently, fairness is not a relevant problem in terms of computer engineering. In non-commercial networks, however, monetary solutions are not available, by definition. The present non-commercial blockchain networks (e.g. test networks such as Ropsten or Rinkeby, or academic networks such as Bloxberg) employ trivial distribution mechanisms called \textit{faucets}, which offer fixed amounts of free tokens (called cryptocurrencies) specific to the given network. This mechanism, although simple and efficient, is prone to denial of service (DoS) attacks and cannot address the \textit{fairness} problem. In the present dissertation, the faucet mechanism is adapted for fair distribution, in line with \textit{Max-min Fairness} scheme. In total, we contributed 6 distinct Max-min Fair algorithms as efficient blockchain faucets. The algorithms we contribute are \textit{resistant} to DoS attacks, \textit{low-cost} in terms of blockchain computation economics, and they also allow for different user weighting policies. While 4 of the contributed algorithms provide \textit{scalability} to unlimited number of users, 2 of them account for both \textit{short term} and \textit{long term} \textit{fairness}.
\end{abstract}

\begin{ozet}
Bu çalışma, ticari olmayan blokzinciri ağlarında, paylaşımlı kaynakların \textit{adil dağıtımı} sorununa hitap etmektedir. Blokzincirleri, belirli bir kullanıcı ağının kayıtlarını kamusal, kriptografik olarak güvenli ve uzlaşımsal yolla sıralamaya ve zaman etiketi vermeye yarayan dağıtık sistemlerdir. Olay, işlem emri, yapılandırılmış işlem kuralları vb. cinsinden olabilen bu kayıtlar, \textit{blok} denilen iyi-tanımlanmış veriyapıları içerisine konur ve bu bloklar kriptografik göstergeler aracılığıyla, zamansal öncelik-sonralık ilikşisini temsil eden bir tümel sıralama içerisinde birbirlerine bağlanır. Blokzincirini işletmek ve/ya bir blokun içeriğine kayıt eklemek, blokzinciri sistemlerinin paylaşımlı kaynaklarıdır. Ticari ağlarda bu kaynaklar itibari para birimleri karşılığında alınıp satılabilmektedir, dolayısıyla bu ağlarda \textit{dağı-lımın adaleti} bilgisayar mühendisliği alanı içinde tanımlanabilecek bir problem değildir. Öte yandan, tanımı gereği, ticari olmayan ağlarda parasal çözümler kullanılamaz. Mevcut ticari olmayan blokzinciri ağları (örn. Ropsten ya da Rinkeby gibi deneme ağları) bu dağıtımı \textit{musluk} adı verilen basit mekanizmalarla sağlamaktadır. Musluklar, belirli bir sayıda, ağa özgü (\textit{kriptoparabirimi} denen) \textit{andaç}ları kullanıcılara bedelsiz dağıtan mekanizmalardır. Basit ve etkili mekanizmalar olmakla birlikte, musluklar hizmet dışı bırakma (DoS) saldırılarına açıktırlar ve adalet sorununa hitap etmemektedirler. Mevcut tezde musluk mekanizması, dağıtımın adaleti sorununa da hitap edecek şekilde Max-min adalet şeması uyarınca adapte edilmiştir. Toplamda, Max-min adalet şemasına uygun ve birbirinden farklı algoritmalarla çalışan 6 adet blokzinciri musluğu, literatüre katkı olarak sunulmuştur. Sunulan bu algoritmalar hizmet dışı bırakma saldırılarına karşı \textit{dirençli} ve blokzinciri hesaplama ekonomisi uyarınca \textit{düşük maliyetli} olmakla beraber, farklı kullanıcı ağırlıklandırma politikalarına elve-rişlidir. Sunulan algoritmalardan dördü sınırsız sayıda kullanıcı için destek sağlarken, ikisi kısa dönem yanı sıra uzun dönem adaleti gereksinimlerine de cevap verebilmektedir.
\end{ozet}
\tableofcontents
\listoffigures
\listoftables
\begin{symbols}
%
\sym{$b_u$}{Resource balance of user $u$, $b_u \in \mathbb{N}, u \in U $}
\\
\sym{$BlockNumber$}{Current block number}
\\
\sym{$c$}{The available capacity, initially $0$, incremented by $C$ every epoch} 
\\
\sym{$Capacity$}{see $c$}
\\
\sym{$C$}{Amount of resource that is added to $c$ at every epoch, $C \in \mathbb{Z}^+$}
\\
\sym{$dt_u$}{Total demand volume of user $u$, including by then present demand}
\\
\sym{$d_{ui}$}{Demand of user $u$ stored on heap $D_i$}  
\\
\sym{$D_i$}{Demand heap $i, i \in \{0,1\}$}
\\
\sym{$Epoch$}{Epoch number}
\\
\sym{$EpochCapacity$}{see $C$}
\\
\sym{$EpochSpan$}{Number of blocks in an epoch}
\\
\sym{$i$}{Index variable}
\\
\sym{$n$}{Number of users, $n = |U|$}
\\
\sym{$Offset$}{The block number at which the contract was deployed}
\\
\sym{$p$}{Decimal precision, internally kept for float variables}
\\
\sym{$ResetEpoch$}{The epoch at which the total weight was last reset}
\\
\sym{$Round$}{Round number}
\\
\sym{$RoundSpan$}{Number of blocks in a round}
\\
\sym{$s$}{Unit share}  
\\
\sym{$s_u$}{User share of user $u$}
\\
\sym{$selector$}{Variable for pointing out the active buffer in circular buffers $selector \in \{0,1\}$}
\\
\sym{$TotalWeight[selector]$}{Total weight for even and odd epochs}
\\
\sym{$u$}{User index variable, id, $u \in [1, n] $}  
\\
\sym{$U$}{Set of users $U = \{ u_1,\ldots,u_n \}$}
\\
\sym{$User$}{User object, $User \in U$}
\\
\sym{$User.balance$}{Resource balance of user $u$, $User.balance \in \mathbb{N^+}$}
\\
\sym{$User.claimEpoch$}{The last epoch user $u$ made a claim}
\\
\sym{$User.claimRound$}{The last round user $u$ made a claim}
\\
\sym{$User.demand[selector]$}{Demand of user $u$ in buffer $selector$}
\\
\sym{$User.dem..Epoch[sel..]$}{The last epoch user $u$ made a demand, $selector \in \{0,1\}$}
\\
\sym{$User.weight$}{Weight of user $User$}
\\
\sym{$Volume$}{Demand volume}
\end{symbols}

\begin{abbreviations}
\sym{AMF}{Autonomous Max-min Fairness}
\sym{CMF}{Conventional Max-min Fairness}
\sym{MF}{Max-min Fairness}
\sym{QMF}{Quantized Max-min Fairness}
\sym{p2p}{Peer-to-Peer}
\sym{PGP}{Pretty Good Privacy}
\sym{PoA}{Proof of Authority}
\sym{PoS}{Proof of Stake}
\sym{PoW}{Proof of Work}
\sym{SMF}{Simulated Max-min Fairness}
\sym{WAMF}{Weighted Autonomous Max-min Fairness}
\sym{WCMF}{Weighted Conventional Max-min Fairness}
\sym{WMF}{Weighted Max-min Fairness}
\sym{WQMF}{Weighted Quantized Max-min Fairness}
\sym{WSMF}{Weighted Simulated Max-min Fairness}
\end{abbreviations}

\chapter{INTRODUCTION}
\label{introduction}
\pagenumbering{arabic}
\renewcommand{\thefootnote}{\arabic{footnote}}

In 2008 an anonymous author, or a group of authors, published a whitepaper under the pseudonym ‘‘Satoshi Nakamoto" describing a distributed system of digital money, named by the author(s) Bitcoin \cite{nakamoto2008bitcoin}. For the first few years it was of interest only to a limited number of technology enthusiasts, futurists, science-fiction fans, and probably to a lesser extent to mathematicians and engineers specialised in the area. After, however, slightly more than a decade now, it is a major financial instrument employed throughout the world. Not only is it a successful financial instrument, its design inspired other systems to come forth, leading to an engineering subfield of its on: \textit{blockchains}. The present dissertation is situated within this subfield.

Departing from a well studied problem in the computer science literature (i.e. distribution of shared resources), we examined the blockchain environments, adapted a conventional solution, Max-min Fairness (MF), to this novel context, and proposed alternatives conditioned on different premises, and serving different use cases. We report experimental results showing that the conventional algorithm cannot be deployed \textit{per se} in the blockchain context. Our adaptations and alterations, on the other hand, work without problems and scale for wide use cases.

The solutions we propose bear relative advantages to each other, and each one is optimal for a different condition. A comparison of Autonomous Max-min Fairness (AMF), Quantized Max-min Fairness (QMF), Simulated Max-min Fairness (SMF) and their weighted counterparts (with leading ‘W's), as we name them, can be seen in Table \ref{comparison}. While AMF is the one with the working principles most similar to the conventional Max-min Fairness algorithm and works on no restrictions, the remaining two (and their weighted versions, likewise) bear an advantage of completing the distribution in fewer transactions, in return for certain restrictions. While QMF operates under the assumption of accepting demand volumes only from a predefined numeric interval, the number of users SMF can support is limited.

\begin{table}[ht]
    \centering
    \caption{Comparison of Contributed Algorithms}
    \label{comparison}
    \begin{tabular}{r|c|c|c}
          &  Claim Rounds & Demand Volume & Number of Users \\
    \hline \hline
    W/AMF & Multiple & Unrestricted & Unrestricted \\
    W/QMF & Single   & Quantized    & Unrestricted \\
    W/SMF & Single   & Unrestricted & Limited \\
    \hline
    \end{tabular}
\end{table}

\section{Blockchain Mechanics}

Although it is possible to review blockchains from various differing standpoints, the present dissertation takes a rather technical/engineering point of view. According to this, a blockchain is a distributed datastructure, which may be used in various different contexts to serve a number of diverse functions. As the name implies, it is a \textit{chain} of \textit{blocks} bound to each other by a specific mathematical method, called \textit{proofing}. These proofs, which may be of a number of different kinds (e.g. Proof-of-Work, Proof-of-Stake, Proof-of-Authority), serve as pointers among the blocks, as well as serving the purpose of establishing consensus among the users for the content of the next block to be appended to the chain in the course of its growth.

\begin{figure}
	\includegraphics[width=\textwidth]{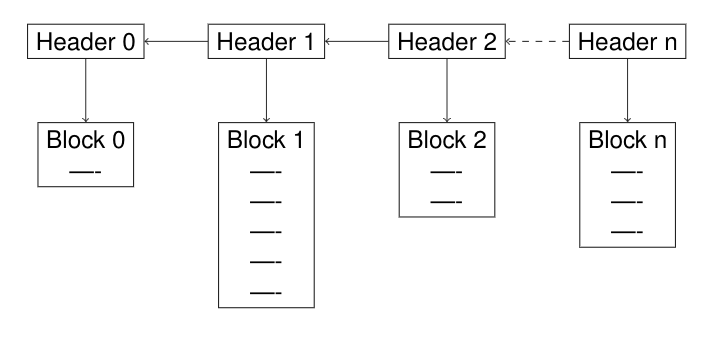}
	\caption{Blockchain Diagram}
	\label{blockchain}
\end{figure}

The blocks referred to here are datastructures which organise records generated by the users within the network. It consists of a header describing the metadata (e.g. time of creation, address of the creator), and trailed to that ordered records from the users (see Figure \ref{blockchain}). Each block refers to another unique block as its predecessor in the chain. It is a total ordering in which succession-precedence relations represent the temporal order of the events as they are agreed upon by the user community. Each user keeps a copy of the blockchain in her local host, and this way the content of the blocks in the chain cannot be altered or deleted once they \textit{become a part of the blockchain}\footnote{For a given block to \textit{become a part of the blockchain} is not a straightforward concept. It necessitates a number of \textit{younger} blocks to follow a given block, in order for the latter to be certain to reside in the chain with overwhelming probability. The reader may refer to \cite{nakamoto2008bitcoin} for a detailed analysis.}, which is commonly referred to as the \textit{immutablity} property of the blockchains.

A blockchain is operated on by means of a \textit{blockchain system}, which is a piece of software, a virtual machine developed and maintained collectively by a community of users. A \textit{blockchain ecosystem}, in turn, is a community of users running the same blockchain system on their local host and synchronising with each other over a p2p network. A host with a blockchain system as a p2p client on is called a \textit{node}, and collectively the nodes govern the procession of the overall system.

Although differing among blockchain systems, nodes share certain basic functionalities such as:

\begin{itemize}

\item listening to broadcasts from the network
\item obtaining the newly appended blocks and checking their proof
\item generating requests from the user and broadcasting them to the network
\item obtaining requests from the user pool and organising them into a block
\item appending a block to the chain by providing a proof
\end{itemize}

As mentioned before, proofing is the main method used in blockchain systems in order to reach consensus. These proofs both secure the identicality of each copy of the blockchain in the network, and the accumulating proofs provide means for future operations on the blockchain. This is achieved by utilising a charging system for operating on the blockchain. When we indicate \textit{operating on the blockchain}, we refer to the transitions on the networkwide agreed upon state of the blockchain system according to the content of the transactions within a block, which is invoked by a local node broadcasting the proofed block to be appended to the chain, and as a result of which each node organises its local blockchain and local state variables to agree with the most up-to-date global state. Not unexpectedly, transaction requests from the network are submitted \textit{in return for} the native accounting unit, which is called a \textit{cryptocurrency}, where appending a new block to the chain by producing a proof is \textit{rewarded} by the same cryptocurrency. As such, blockchains are \textit{incentive driven} systems. Their operation is dependent on perpetual user involvement and contribution.

In addition to the above-described basic functionality that is common to all blockchain systems, starting with Ethereum \cite{wood2014ethereum} and attaining wide acceptance among different communities, today majority of the blockchains offer Turing Complete functionality over the scripts that can be included in blocks and interacted with. Ethereum Virtual Machine (EVM) and the \textit{Solidity} scripting language have reached wide popularity in the cornucopia of blockchain system designs of the last decade and became the \textit{de facto} standard \cite{hildenbrandt2018kevm}. They are also employed in the experimental setting of the present dissertation.

Since Turing Machines are subject to the halting problem, and since programming errors may lead to infinite loops, both of which are serious threats for the sustainability of distributed systems, a safeguard mechanism is built into the operation of EVM. This mechanism, called the \textit{block gas limit}, sets an absolute upper bound to the number of operations that may take place within the execution of a single block.

As described above, each blockchain operation is charged in return for the blockchain's native cryptocurrency. In the context of EVM smart contracts, each assembly level operation is charged by a predefined amount, the unit of which is referred to as \textit{gas} and exchanged in return for cryptocurrency. The total cost of any given function in units of gas, may not exceed the block gas limit, for it to be executed. If the execution of a given function causes EVM to reach block gas limit, this is detected in runtime and the system state is rewinded back to the point where the function started executing. The function exits, returning an error message to the caller.

In commercial networks, the cryptocurrency is either \textit{mined} by providing a proof and contributing a block to the system, or purchased in return for fiat currency. This bears a problem for testing software on the network before deploying it, for it renders the development process unnecessarily costly. For this purpose test networks have been designed and deployed, and they also are in wide use recently. These networks employ identical mechanisms to the commercial blockchain networks, with the exception of offering free cryptocurrency to its users. Consequently, these are non-commercial networks, and they use an alternative cryptocurrency distribution mechanism called a \textit{faucet}, which offers a fixed amount of the network's native cryptocurrency to any demanding user. The user, in turn, can deploy her contract on the test network to test it, and the total gas cost of the contract, converted to the cryptocurrency, is charged by the network over the sum obtained from the faucet. Likewise, the user may interact with her contract to test the functionality by sending transactions to the network, the gas cost of which are, again, converted to the cryptocurrency and charged over the sum obtained from the faucet. In case of depleting the sum obtained from the faucet, the user may send new requests to the faucet to obtain the fixed amount one more time. The user may repeat obtaining cryptocurrency from the faucet for an unlimited number of times.

As can be easily perceived, this system is simple and efficient, but it can hardly be dependable for securing \textit{fairness} of distribution among its users. Any adversarial party may exploit the system simply by submitting recurrent requests to the faucet and accumulating the obtained cryptocurrency, with which she can launch Denial of Service (DOS) attacks. The present dissertation focuses on this problem and offers solutions.

\section{Max-min Fairness Distribution Scheme}
\label{mm_model}

The main objective of the MF scheme is to maximise the minimum share given to any user. Although it is possible to define MF also for continuous time (e.g. for queueing fairness for network flows), blockchains are implicitly discrete time systems, thus we will suffice describing it in discrete time. The reason for blockchains being discrete time systems is that all events and state transitions occur as a result of a new block being appended to the chain, which is mutually exclusive among candidate blocks. In other words, since by definition only one block may be appended to the \textit{chain} in unit time and in the time between no event or state transition can take place, the system time is well dissected and discretised.

The procession of MF is based on a trivial fairness scheme, where resources are uniformly distributed among the demanders, each one of the $n$ demanders obtaining $\frac{1}{n}$ of the resource. MF improves the trivial scheme on the premise that not every demander would demand as much as the share that is reserved for her. Accordingly, the MF allocation algorithm takes recursive iterations over the list of demanders, reallocating unused shares of the underdemanders among the overdemanders.

In the first iteration, starting with the smallest demand and proceeding in the ascending order, the algorithm allocates the demanders the minimum of $\frac{1}{n}$ of the capacity ($c$) and their demands (i.e. $\min\left(\frac{c}{n}, d_u\right)$). At the end of the first iteration, some demands are fully supplied and removed from the list of demands, and some capacity is left over. The algorithm, in turn, proceeds with updated $n'$ and $c'$, until either all demands are fully supplied, or the capacity is depleted.

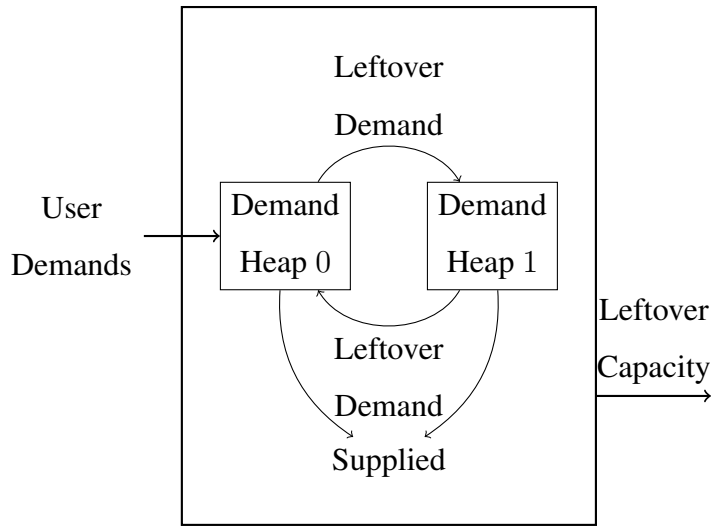
\begin{figure}
	\centering
		\begin{tikzpicture}
        \node [draw,rectangle,align=center] (dho) {Demand\\Heap $0$};
        \node [draw,rectangle,right = of dho,align=center] (dhi) {Demand\\Heap $1$};
        \node (sup) at ($(dho)!.5!(dhi)+(0,-3)$) {Supplied};
        
        \draw [->] ($(dho.north)!.5!(dho.north east)$) to [bend left, above, out=60, in=120] node [align=center] {Leftover\\Demand} ($(dhi.north west)!.5!(dhi.north)$);
        \draw [<-] ($(dho.south)!.5!(dho.south east)$) to [bend left, below, out=-60, in=-120] node [align=center] {Leftover\\Demand} ($(dhi.south west)!.5!(dhi.south)$);
        
        \draw [->] (dho) to [bend right] (sup);
        \draw [->] (dhi) to [bend left] (sup);
        
        \draw[thick] ($(current bounding box.south west) - (.5,.5)$) rectangle ($(current bounding box.north east) + (.5,.5)$);
        
        \draw [<-,thick] (dho.west) --++ (-1,0) node [left,align=center] {User\\Demands};
        \draw [->,thick] ($(current bounding box.east)!.5!(current bounding box.south east)$) --++ (1.5,0) node [midway,above,align=center] {Leftover\\Capacity};
		\end{tikzpicture}
    \caption{The operation of Max-min Fairness Algorithm}
	\label{max-min_diagram}
\end{figure}

The operation of the MF scheme is schematically represented in Figure \ref{max-min_diagram}, and its pseudo-code is presented in Algorithm \ref{mf_pseudo}. In the pseudo-code the demand heaps are denoted by $D_0$ and $D_1$, and individual demands in these heaps are represented by lower case letters, subscripted with $u$, for user id number (i.e. unique identifier given to each user, denoted $d_u$).

The balances of users are kept in a vector, and the balance of user $u$ is represented with $b_u$. At each iteration, the maximum available amount to be allocated to each user is recalculated by dividing the remaining capacity by the number of remaining demands, and denoted by $s$, representing the \textit{unit share}.

To illustrate the operation of the algorithm we may consider the following example: Suppose that a resource of $30$ units will be shared among three users, with the demands submitted as $<$$4,11,15$$>$. The algorithm distributes the resource in $3$ iterations. The rounds and the shares assigned in each round can be seen in Table \ref{max-min_example}.

\begin{table}
    \caption{An exemplary distribution according to Max-min Fairness scheme}
    \label{max-min_example}
    \centering
    \begin{tabular}{r|rrr|rr}
         \multicolumn{1}{r}{}  & User $1$   & User $2$   & User $3$  & Share    & Capacity  \\
         \hline\hline
         Demands        & $4$   & $11$  & $15$  &           & $30$      \\
         \hline
         Iteration $1$  & $4$   & $10$  & $10$  & $10$      & $6$       \\
         \hline
         Iteration $2$  & $0$   & $1$   & $3$   & $3$       & $2$       \\
         \hline
         Iteration $3$  & $0$   & $0$   & $2$   & $2$       & $0$       \\
         \hline
         Total          & $4$   & $11$  & $15$  &           &           \\
         \hline
    \end{tabular}
\end{table}

How the unsatisfied demand, or the leftover capacity will be treated after a distribution period is a decision of \textit{policy}. In our current work, we implement a policy that discards all the unsatisfied demands, in the case of capacity depletion, and hands the leftover capacity over to the next distribution period, in the case of satisfying all the demands.

\begin{algorithm}[t]
	\begin{algorithmic}[1]
\Procedure{Distribute}{$DemandHeap, Capacity$}\Comment{Distribute Centrally}
\State $Capacity \gets EpochCapacity$;
\State $selector \gets 0$;
\While{$DemandHeap[selector].size > 0$ \textbf{and} $Capacity > 0$}
	\If{$Capacity < DemandHeap[selector].size$}
		\State $Share \gets 1$;
	\Else
		\State $Share \gets \floor*{ \frac{Capacity}{DemandHeap[selector].size} }$;
	\EndIf
	\While{$DemandHeap[selector].size > 0$ \textbf{and} $Capacity > 0$}
		\State $User.balance \gets User.balance + \min{(Share, User.demand)}$;
		\State $Capacity \gets Capacity - Volume$;
		\If{$User.demand > Share$}
			\State $DemandHeap[1 - selector].insert(User.demand - Share, u)$;
		\EndIf
	\EndWhile	
	\State $selector \gets 1 - selector$;
\EndWhile
\EndProcedure
\end{algorithmic}

	\caption{MF Pseudocode}
	\label{mf_pseudo}
\end{algorithm}

The amount that is reserved for each epoch is denoted by $C$. We call this amount the \textit{epoch capacity}, and in the present dissertation, we took it to be constant. The actual amount that is distributed in an epoch is denoted by $c$, and it is at least as much as $C$, since it is added to $c$ at the beginning of each epoch (i.e. Algorithm \ref{mf_pseudo} line $2$).

In Algorithm \ref{mf_pseudo}, the lines $4-19$ constitute the main, or outer loop of the algorithm, which is responsible for repeating the inner loop (lines $10-18$) until either the demands or the capacity is depleted. It starts with calculating the initial share (lines $5-9$), and then starts the inner loop. Once the proceeding of the inner loop is completed, the demand heaps exchange their functions (line $19$) and the outer loop takes another iteration.

The inner loop accounts for iterating on and processing the demands in the \textit{active} heap. In line $11$ the demand volume and the user id at the root of the heap is read into a variable and deleted from the heap. After that the minimum of the user demand and the unit share (i.e. $\min\left(\frac{c}{n}, d_u\right)$) is assigned to the user in lines $12-14$. The control structure in lines $15-17$ checks whether the demand is fully satisfied or not. If not, the leftover demand is inserted to the \textit{passive} heap with the user's id (line $16$) to be processed in further iterations.

Another version of MF is \textit{weighted MF}, in which case the users are weighted over some predefined policy, and the shares are calculated with the weights assigned to each user, individually. In this version, instead of the number of demands, the total capacity is divided by the \textit{total weight} in order to calculate the unit share ($s$). In turn, the \textit{user share} ($s_u$ for user $u$) is calculated for each user by multiplying the unit share with the user's weight ($w_u$ for user $u$). The users are allocated the minimum of their demands, and their individually assigned user shares. Accordingly, the formula for calculating the unit share $s$ is:

\[s = \frac{c}{\sum_{u=1}^{n} w_u}\]
\noindent
and the user share $s_u$ is given by:

\[s_u = w_u \cdot s = w_u \cdot \frac{c}{\sum_{u=1}^{n} w_u}\]

\section{Contributions to the Literature}

Distribution of shared resources is a generic problem that we encounter as the distribution of cryptocurrency, in the blockchain context. As the use cases of blockchains grow, trading of these resources in return for fiat currency remains a limited solution which cannot be applied to non-commercial blockchains. For this reason, algorithmic, fair, and secure distribution of these resources acquire importance in the process. The solutions offered by the present dissertation make the following contributions:

\begin{enumerate}
    \item Max-min Fairness, a well known and well studied fair distribution scheme, is actualised in the blockchain context as smart contracts, to avail a generic solution for diverse use cases. The algorithms developed in the present dissertation are open to public use at \cite{metin2020faucet} to be adapted or modified easily for the needs of the projects that intend to use them. These include at first hand the test networks such as Ropsten and Rinkeby, and academic networks such as Bloxberg \cite{bloxberg}.
    \item Four algorithms AMF, WAMF, QMF and WQMF operate independent of the number of users enrolled in the system, therefore they can scale up to unlimited numbers of users.
    \item Two algorithms, SMF and WSMF, can scale up to $250$ users without running into the block gas limit\footnote{In the present dissertation the block gas limit is assumed to be 8.000.000, which was the actual number as of time the tests were being carried out.} exhaustion problem, which can be used by small-to-mid-size communities. Although not as general a use case as being scalable to unlimited number of users, SMF and WSMF offer other advantages for communities conforming to the allowed size.
    \item All of the functions implemented in the smart contracts are cost-efficient in terms of gas expenditure. Majority of them are constant or near constant (i.e. negligibly variable) cost. The only functions that are significantly dependent on the input size are the share calculating functions of SMF and WSMF, which are conditioned on the number of users, hence the limitation.
    \item Weighting policies defined for WAMF and WSMF can account for \textit{long term fairness}, in addition to the short term fairness intrinsic to the Max-min distribution scheme. Unfortunately, the same policy is not applicable to WQMF for reasons that will be discussed in Section \ref{wqmf_model}.
\end{enumerate}

In addition to the main objectives of the dissertation, some minor contributions may be counted as follows:

\begin{enumerate}
    \item Over the implementation of CMF, it has been shown that a conventional approach for implementing Max-min Fairness scheme, in which a central executive unit does the calculation and assignment of the shares, is not efficient in blockchain context. This is a general result, demonstrating the fact that in the blockchain programming context, the loops in any algorithm should preferably be distributed over the users to be executed decentrally.
    \item An array implementation of minimum heap is contributed, since Solidity does not offer a heap datastructure. The algorithm can be accessed at the repository \cite{metin2020faucet}.
    \item Although AMF is an adaptation of conventional Max-min fairness algorithm to the blockchain context, simply distributing the main loops over the users and the claim rounds, to the best of our knowledge, the structure of QMF and SMF are novel not only in blockchain context but also as stand-alone algorithms themselves. Instead of gradually assigning shares to users in multiple iterations as it is done in the conventional algorithm, QMF and SMF calculate the maximum share that may be offered to users, such that when the user takes the minimum of the available share and her demand, the total capacity that is available to be distributed is not exceeded.
\end{enumerate}

\clearpage

\section{Dissertation Outline}

The rest of this dissertation is organised as follows: 

In Chapter \ref{related_work}, we review the literature on blockchain systems, fairness assumptions of various proof schemes, and the problem of fair resource distribution. Although it is possible to handle it from a diverse number of points of view, we keep the discussion of the proof systems limited to their relation to the resource distribution process, and touch upon the problems of distributed trust and distributed consensus to the extent which it relates to our discussion. It should be noted that a comprehensive discussion of proof systems necessitates a larger context, yet it is beyond the scope of the present dissertation.

Following that, in Chapter \ref{problem_statement}, we lay down our main constructs and formulate the problem at hand in terms of them. The main point of this chapter is the justification of the main metric employed in our tests, since it is rather inconventional for the computer science literature. In the same chapter we also describe the testing environment and we formulate the problems related to the temporal setting. Timing and synchronisation is a general problem intrinsic to all distributed systems, and in blockchain systems these problems remain, taking a specific form. We describe this form and define our solutions to these problems in this chapter.

In Chapter \ref{max-min_on_blockchain}, we compare the conventional implementation of the algorithm (CMF) with its autonomous counterpart (AMF) and show that while the former does not scale for even small number of users (i.e. $\sim10$), the latter does scale for unlimited number of users. We also lay down the experimental setting in this chapter, which will be used also in the following chapters for testing the remaining algorithms, although with small modifications.
 
In Chapter \ref{max-min_restructured}, we present the restructured versions of the algorithm. We begin the chapter by describing the models QMF and SMF in their abstract forms, and give the implementation details in the subsequent sections. After indicating the small alterations in the experimental setting from the previous chapter on AMF and laying down the new parameters, we present the experimental results of W/QMF and W/SMF in comparison to both each other and also to W/AMF.

In Chapter \ref{discussion}, we present a discussion on the findings of the present study and propose alternative points of view on the problem.
 
We conclude the dissertation in Chapter \ref{conclusion} with suggestions for prospective studies.

\chapter{RELATED WORK}
\label{related_work}

The first blockchain, Bitcoin \cite{nakamoto2008bitcoin} has been developped as digital money, thus the fair distribution of resources in this context means the fair distribution of the total reserve of coins in the ecosystem. Bitcoin network started with no initial reserve, and it creates new coins as a reward for each block newly added to the chain. In its first years, no user participated in the system, therefore Satoshi Nakamoto mined blocks that has single transactions, transferring the newly mined coins to the miner of the block. In block $170$, Nakamoto made the first p2p transaction by sending $10$ Bitcoins to Hal Finney, a renowned cryptographer.

As its precursors \cite{dai1998b, back2002hashcash} Bitcoin utilises Proof-of-Work (PoW) as its proof scheme. In order to append a newly created block to the chain, the users are expected to find an input to a \textit{one-way function} that results in a targeted output. Since the function is one-way, meaning, it is not possible to calculate the input that results in a given output, the only way to find such an input is by trial-and-error. The user creates the block according to the predefined block syntax. After that she iteratively appends \textit{nonces}, calculates the images of the resulting strings under the one-way function, and checks the outputs against the properties of the targeted structure, until she \textit{comes upon} a desired output. Because of the fact that the user randomly \textit{explores} the output space by systematically trying inputs from the input space to come upon a desired output, the process is referred to as \textit{mining} a block, and the user is referred to as a \textit{miner}.

Although the one-way function, which is sometimes called a \textit{puzzle}, may be of various types, the most commonly used ones are cryptographic hash functions\footnote{In the case of Bitcoin, for example, this is SHA256 \cite{nist2015180}.}. A cryptographic hash function guarantees that the probability of each output string coming up for a given input string is equal. The targeted output structure provides a restriction on the image set of the hash function for strings to be accepted. Thus, the probability of finding a proof for a given string is the ratio of the size of the restricted image set to the size of the total image set. In Hashcash, for example, the target is defined by the number of leading $0$'s in the output, in Bitcoin the target is defined by the output being smaller than a decided number. Greater the number of leading $0$'s in the former, or the smaller the decided number in the latter, smaller the set of accepted strings, thus smaller the propability of coming upon a number in this set. This is referred to as the \textit{difficulty} of the puzzle.

The difficulty parameter is a moving average, updated periodically to keep the time to append a new block, referred to as \textit{block latency}, constant \cite{nakamoto2008bitcoin}. In Bitcoin, for example, the difficulty updates are done every 2016 blocks to keep the block latency at approximately $10$ minutes. This update is implemented as a function hardcoded into the blockchain system, which is called every 2016 blocks. The function checks the time it took to produce last 2016 blocks, from this and the current dificulty, it estimates the total processing power utilised by the ecosystem, and calculates the new difficulty needed to keep the block latency at $10$ minutes with the present total processing power \cite{garay2015bitcoin}. Therefore, as new miners are involved in the system and the total processing power increases, the difficulty also increases to match the new total and keep the targeted block latency in check. Similarly, if miners leave the system, the difficulty decreases.

The probability of finding a proof before the other users is proportional to the number of inputs the user can try in unit time with the processing power she utilises. This way, among its various functions, PoW also accounts for the fairness of the system, because any willing user can participate in the mining process, and obtain coins proportional to the computing resources she is willing to contribute to the system. The users that are not willing to participate in the mining process can purchase coins from the coin owners, which is subject to the market dynamics of demand and supply, which also is widely considered fair.

Although fairness is intrinsic to PoW systems by the means described, these systems expend enormous physical resources. According to \cite{de2018bitcoin}, the energy consumption of Bitcoin network is comparable to the energy consumption of Austria or Ireland. This motivated researchers to seek for alternative proof systems. One such system that has been developed through extensive discussions in the Bitcoin forum is Proof-of-Stake (PoS) which relies on the total volume of coins a user retains in the ecosystem, in other words the user's \textit{stake} in the ecosystem, to decide on the party to append a new block and obtain the block reward. This process is referred to as leader election.

By definition, PoS systems need a method for initial distribution of coins, since the creation of new blocks depend on users who \textit{own} coins, referred in this context as the \textit{stakeholders}. One way is to resort to extra-digital methods, such as uniform or arbitrary distribution among previosly known and trusted users, or \textit{initial coin offerings} (ICOs) and \textit{airdrops} \cite{li2021operation, froewis2021rise}. An ICO is the process of selling coins (typically in an auction) prior to the deployment of the blockchain, and an airdrop is the process of giving away coins for free to the parties that the developers aim to incentiveise joining in the ecosystem. The fairness of the protocols utilising these methods relies heavily on the fairnesss of the initial distribution process which cannot be algorithmically accounted for.

Algorithmic accounts for the fairness of the PoS systems are mainly concerned with the randomisation\footnote{It should be noted that \textit{secure} and \textit{verifiable} randomisation in distributed systems with untrusted parties is not a trivial task. See, for example \cite{de2021randsolomon}.} of the leader election process \cite{bentov2016snow, kiayias2017ouroboros}, assuming that there is an intial set of stakeholders, and relying on this set for launching the system is fair. According to this, if the leader election is verifiably randomised in a weighted way, and the weights are assigned in proportion to the stakes, the block mining rewards are distributed equitably among the users, proportional to the stake they hold.

As has been indicated, PoW blockchains are able to operate in \textit{trustless} networks, meaning, the parties do not need to know and trust each other, in order to trust the security and fairness of the system. That is because the structure of the proof system itself \textit{generates} trust for the ecosystem, the members of which are unknown to each other. PoS systems, on the other hand, assume some degree of trust, at least in the initial deployment of the system, to a subset of parties. On the other extreme reside the Proof-of-Authority (PoA) systems, which need a set of users, referred to as the \textit{authority nodes}, unconditionally trusted by the members of the ecosystem, since they hold the exclusive right to create and append blocks to the chain.

PoA is a natural idea for computer scientists, since its trust structure  is identical with that of the conventional computer environments, where there is a \textit{server} serving \textit{clients}. Formally, this architecture is known as the \textit{client-server architecture}, and the underlying mechanism for the parties to securely identify each other is known as the Pretty-Good-Privacy (PGP) trust model. In PGP there is a party called a \textit{trust anchor} which is trusted unconditionally, and the other parties are authenticated either by the direct reference of this trust anchor, or by a chain of references rooted at it \cite{abdul1997pgp}. In the case of internet, for example, this trust anchor is user's Domain Name Server (DNS) resolver, which takes queries from the user and starts the query chain that reaches the \textit{root nameserver}, and the answer to the query returns to the user through same chain. The user trusts the answer because she trusts her resolver, and in turn each node in the chain trusts the node it queries. Similarly the users in a PoA ecosystem trust the authority nodes, and the rest of the ecosystem, and the operations are trusted over their authenticity and authentication. If authority nodes behave unfairly, for example selectively accept or order transactions to favour a subset of users, authenticate illegal transactions, or even stop the working of the blockchain altogether, there is no way for the users to check on it.

There are also hybrid proof systems that alter and combine the working principles of these proof systems. In fact, the first PoS chain, Peercoin is one such system, initialised with PoW to handle the initial distribution process, and then shifted to PoS as the difficulty of the mining process increased to a certain level in time \cite{ge2022survey}. The fairness of distribution in these systems revolve around the same tenets as described for PoW, PoS, and PoA systems, with different proportions of mixing from one or the other. Some of these are Proof-of-Prestige \cite{krol2019proof}, Proof-of-Activity \cite{bentov2014proof}, Proof-of-Useful-Work \cite{ball2017proofs}, among others.

Before proceeding to the systems built on top of blockchains, a disclaimer is in place at this point. Not all proof systems mentioned here are justified for their accomplishment of fairness or provision of trust. As such, we do not endorse or vouch for the reliability or validity of the claims of these systems. The aim of this literature review is to present the arguments of existing systems, which in reality are subject to the testing of history and human experience. We are well aware that this is not common in engineering practice, but blockchain systems are a subfield of economy and social sciences, as much as they are of engineering, because of their intense entanglement in financial constructs.

\clearpage

The second generation blockchains, starting with Ethereum \cite{wood2014ethereum}, are charactarised by \textit{smart contracts}, which are scripts that reside on the blockchain and interpreted by the blockchain's virtual machine. Ethereum Virtual Machine (EVM) is \textit{Turing Complete}, meaning, it can carry out any calculation that can be carrired out with a Turing Machine. Smart contracts can define, store, and manipulate data, thus with their introduction \textit{second layer coins} logically ensued.

A second layer coin is basically a smart contract that operates according to the mechanics of a cryptocurrency. As such, they inherit the strengths and the vulnerabilities of the blockchain they reside on, and build their operation on top of them. Although designing and developing a coin in a smart contract is easier as compared to doing it within the working of a proof scheme, since the latter is further burdened with addressing the other needs of a blockchain system such as providing digital trust, distributed consensus, ordering and synchronisation etc., fairness of distribution remains a problem to be solved for these coins too. 

Another variety of a second layer coin is a \textit{token}, which is basically a coin, representing a \textit{specific} kind of resource, as opposed to the \textit{generic} nature of the coins. For example \textit{governance tokens} are used as a means of exchange for voting rights in collective decision making for the governance of blockchains, or other decentralised exchanges \cite{froewis2021rise}. Today, there are two token standards available and in general use. These standards are defined by \textit{Ethereum Request for Comment} (ERC) documents, the function and the structure of which are inspired by \textit{Request for Comment} (RFC) documents. The token standards defined in ERC20 and ERC721 define divisible and non-divisible, or \textit{fungible} and \textit{non-fungible} tokens, respectively \cite{shirole2020cryptocurrency, di2020tokens}. Nevertheless, a token may also be issued as a coin, simply by depriving it a specific context. Majority of the existing second layer coins use ERC20 standard. Similarly in common use, ERC721 is generally used to represent digital objects, such as digital art pieces, in-game items etc. \cite{casale2021networks}.

The disentanglement of coin mechanics with the substructural needs of a blockchain ecosystem, and the introduction of tokens enable us to handle the problem of fair distribution in isolation, and adapt the solutions developed for traditional problems. The question of fair distribution first arose in the context of operating systems, where scheduling the resources of a single computer (e.g. processor time) among \textit{processes}, typically at the computer centres of universities, was the main problem \cite{waldspurger1995lottery}. Although it is a fair policy to distribute the resources among the processes, it is prone to degeneration by adverserial users, simply by dividing a task into multiple processes. This lead administrators to implement policies distributing the same resources among \textit{users} \cite{kay1988fair}, and/or \textit{user groups} \cite{mohanty2020qos}.

Similar problems are addressed in the computer networks literature over the allocation of link bandwidth \cite{nace2008max,hahne1991round}. Fair scheduling algorithms have also been the focus of attention in grids \cite{doulamis2007fair}. With the advancements in distributed systems, and new paradigms in cluster and high-performance computing, the problem of fairness evolved yet to larger scales, and new questions arose. In this context, typically, service providers charge users for the common resource that is demanded by, and allocated to them. The same question is now expressed in terms of charging fairness: how much should each demand cost, for it to be fair among clients \cite{marbach2002priority}? Should each type of resource cost the same, and if not how are they traded \cite{ghodsi2011dominant}?

In many areas in computer science where the problem of distributing shared resources is encountered, Max-min Fairness  \cite{bertsekas1992data, keshav1997engineering} has been considered as a fair method \cite{hahne1991round, gogulan2012max}. It is also the main method employed in the present dissertation.

Blockchain systems differ from conventional systems for their operation bottleneck, and consequently the algorithms that run on these systems differ for their design and performance analysis. A number of studies have been offered for the evaluation of performance \cite{alharby2019blocksim}, principles on the algorithm design \cite{marchesi2020design} and the robustness \cite{canfora2020gasmet}. In \cite{alharby2019blocksim}, Alharby et al. develops a simulation environment to evaluate the design and the deployment choices in the development of blockchains. All of these studies concentrate on the block gas limit exhaustion problem, which is a counterpart of and a metric for the \textit{computational cost} of a given algorithm in the conventional setting. Conveniently, the present dissertation uses the same metric for assessing performance.

\chapter{PROBLEM STATEMENT AND TESTING ENVIRONMENT}
\label{problem_statement}

As explained in Chapter \ref{related_work}, among their various functions, proof schemes serve for fairly distributing native coins of the system, to some definition of fairness varying among differing proof schemes. In PoW systems, the fairness is derived on the basis of contributing processing power, in PoS systems over the initial investment and evolving stake of the users, etc. These are all dependent on extra-digital economic systems, such as the investment on hardware in PoW, or investment in fiat currency in PoS. The present dissertation, on the other hand, is focused on handling the same problem in non-commercial ecosystems, such as test networks such as Ropsten or Rinkeby, or scientific networks such as Bloxberg.

In the absence of economic interests and when it is fair to assume the necessary degree of digital trust is provided to the ecosystem by other means, blockchain systems has still much to offer such as transparency, data redundancy, commitment etc. For this reason we handle the fair distribution problem in isolation. To accomplish this we design and test our algorithms on PoA blockchains, which, as explained in Chapter \ref{related_work}, do not run into the overheads caused by the constructs we stated above to leave out. Nevertheless, irrespective of the proof scheme, our results are generalisable to all blockchain systems that is compatible with running scripts, since our solutions reside in the second layer (see Chapter \ref{related_work}).

To repeat from the previous chapter, the main problem in the second layer is to keep the algorithm run under the block gas limit. The charging system for the second layer solutions assign a gas cost to each assembly level operation. For example, in Ethereum this is defined in its white paper \cite{wood2014ethereum}. Although it is subject to minor changes over time, this cost structure is assumably more or less constant in between the operations. What we mean to express is that the cost of operations relative to each other may change in \textit{rate} over time, but a costlier operation tends to remain costlier than a comparatively affordable one. For example, in Ethereum, the cost of reading from a non-volatile memory (referred to as \textit{storage}) register has increased over time, yet it remained lower than writing to it. As such, we postulate that as long as the structure of the virtual machines and programming environments remain the same, the cost structure of the operations will tend to remain the same.

The main reason for this tendency is the fact that, not unreasonably, the gas cost of an operation is determined over its expenditure of the system's resources. For this reason, the costliest operation is writing to storage, since it results in occupying a region in the memory of EVM for the long term, as compared to a reading or arithmetic operation, which is executed in runtime and consumes only the processing power at that instance.

With the same logic, in contrast to the traditional approach, the efficiency of an algorithm is determined over its gas expenditure, and not its algorithmic complexity. Of course, the two concepts are related, and a complex algorithm is likely to expand more gas as compared to a simpler one, but in the traditional algorithm analysis, the complexity is measured over the operation that occurs most frequently, indifferent to the type of operation. In smart contract context however, the type of operation makes a difference. For example, traditionally two algorithms running in constant time with the same constant coefficient is considered equally efficient, but in smart contract context, if one is doing a storage write, and the other is doing arithmetic operations, the latter is accepted to be more efficient.

Accordingly, the present dissertation takes the gas expenditure as the main metric for the efficiency of the algorithms developed hereby. We report absolute gas costs of the functions and how they scale with growing values that are critical to each one. As indicated, the gas cost of each opertion is subject to change over time and these absolute values will most probably be obsoleted in the future. Nevertheless, the structure of gas costs and how they scale with the growth of their critical variables will tend to remain accurate as long as the structure of the development and processing environments remain unaltered.

\section{Testing Environment}

We implemented our algorithms in Solidity programming language and run on an EVM environment \cite{dannen2017introducing}, and more specifically, its Parity implementation, as mentioned above. The main reason for selecting this framework is its wide use among blockchain ecosystems. Many blockchain ecosystems and blockchain based systems utilise either EVM or virtual machines similar to EVM, and support Solidity programming language for smart contracts (e.g. \cite{baird2018hedera, tron, cheng2019ekiden, niloy2021blockchain}), and for this reason there are also studies available on the performance \cite{wohrer2018design}, security \cite{wohrer2018smart}, and inspection \cite{bragagnolo2018smartinspect}. It is a high-level, easy to read, object oriented script language.

We tested our algorithms in a local blockchain, operated by Parity Ethereum $2.7.2$ \cite{parity}, and we implemented the smart contracts in Solidity $0.5.13$. The block gas limit we assume is $8.000.000$ units\footnote{This value is taken from the Ethereum's actual block gas limit at the time we started our experiments. The value is dynamically set and updated in each system by the collective decision of its miners, thus it is not up-to-date. Nevertheless, for the purpose of the present study, it does not constitute a problem, and even an up-to-date value will be obsoleted in the time of the readers view so we left it as it is.}.

Parity implementation of Ethereum offers customisable consensus protocols. Among those is the so-called \textit{instant seal engine}, which places each transaction into an individual block of its own. The engine is specifically designed for contract development, since the block latency is rarely a relevant parameter in the development and verification processes of the algorithms, at least for the time being.

For our case, the instant seal engine also allows us operationalise \textit{time} in terms of number of blocks and, in turn, define the \textit{epoch span} and the \textit{round span} in terms of it. As such, our results are generalisable to every blockchain environment, independent of the consensus algorithms and temporal parameters they employ.

\section{Timing and Synchronisation}
\label{synchronisation}

Timing and ordering of events in distributed systems operating asynchronously and in the absence of a central timestamp server has been a field of study since the emergence of such systems \cite{lamport1978time}. As explained in the original article \cite{nakamoto2008bitcoin}, a crucial function of a blockchain is that it serves as a timestamp server in an environment of parties with conflicting interests. Although there is a timestamp in the header of each block in the chain, it is created by the miner, and in a trustless computation environment it may not be exactly reliable. Moreover, the main communication method in blockchain ecosystems is P2P broadcasting, and consequently the latency of the arrival of a message is not uniform among the nodes. Neither is it uniform accross time, since, for a given node, the proof producing node may be topologically close for one block, but the next one might come from a distant node in the network. The only information that a node can obtain for sure is that the proofed block has been produced \textit{before} it has arrived, as it is called the \textit{happened-before relation} \cite{lamport1978time}. In fact, the block headers are checked also for this relation upon arrival, in order to prevent an adversary party exploit the system by timestamping their proof for a later point in time.

As indicated in Chapter \ref{related_work}, the difficulty of the proof is updated every 2016 blocks in Bitcoin network. This interval of regular difficulty update is commonly referred to as an \textit{epoch}, and is employed in all PoW blockchains. In the process of update, each node takes the time difference between the first and the last blocks of an epoch, and divides it to the number of blocks to obtain an \textit{estimate} for the average time interval between the blocks, or as it is commonly called, \textit{block generation rate} or \textit{latency}. This is also an estimate for the total processing power poured into the system, lower latecy meaning higher processing power. The node then calculates the necessary new difficulty to adjust the network with the up-to-date processing power estimate to conform to the targeted block latency.

Situated as such, a convenient measure of time between two events in a blockchain system is the number of blocks between them, as compared to the metric time. It is arguably more relevant a measure than the metric time also because for most of the calculations the \textit{ordering} of events is the relevant factor for the \textit{correctness} of the calculation. For these reasons, in our experiments we used this metric for measuring time and synchronising nodes. Inspired from difficulty update process, we divided distribution periods into \textit{epochs}, and in W/AMF, we subdivided the epochs into \textit{rounds}, defined likewise.

The main concern for the decision on the span of the epochs and the rounds is that they should last enough for each user to be able to make claims and demands within their due interval. Since the instant seal engine deployed in the tests place each transaction in an individual block, the epoch and round spans are so chosen as to allow each user be able to make enough claims and a demand within an epoch. The spans of these epochs and rounds will be given and justified in their relevant sections, for the tests differ slightly between algorithms, to test different constructs they bear.

Since blockchains are incentive driven systems, it is not possible to spawn a daemon process to keep the global state variables up-to-date. The method for maintaining the global variables in such a system is to build a dedicated function for updating the state and call it at the beginning of each function call. This way the system state is collectively maintained by the user community.

In the present algorithms, \texttt{update\_state} is one such function, which is called at the beginning of all other functions, except an internal function of its own, which is called \texttt{calculate\_share}. The \texttt{update\_state} function checks the block number, and from its distance from the block that the contract is deployed, calculates the epoch and round numbers. If an epoch and/or round update is necessary, the other necessary updates such as replenishing capacity or recalculating share is done along with it. Majority of the state update checks return negative and they do not impose a significant additional gas cost on the calling function. In cases where the check return positive and global variable updates are undertaken, the additional gas cost of \texttt{update\_state} is significant, and for this reason it is calculated within the function and returned to the calling user for reasons of fairness.

\chapter{MAX-MIN FAIRNESS ON BLOCKCHAIN}
\label{max-min_on_blockchain}

We develop autonomous algorithms  AMF and  WAMF for actuating  the MF scheme. In WAMF, the weights are defined to be the reciprocals of the total amount of demands users have made up to the distribution time. This aims at incentivising users to make minimal demands suitable to their needs, in order not to be disadvantageous in the long run. The implementation details of WAMF algorithm, as well as its pseudocode is presented in Section \ref{ca_implementation}.

\section{Implementation}
\label{ca_implementation}

The conventional setting to utilise MF typically includes a central unit (either an individual process running on a central processor or a dedicated administrative host in a computer network) calculating the shares and carrying out the iterative assignments. This is applicable to the blockchain context, but not without potential drawbacks. The main bottleneck in such an adaptation is the block gas limit, which imposes an absolute upper bound for the number of operations that may take place within the processing of a single block. For this reason, we implemented two algorithms and compared them. The implementations are available in \cite{metin2020faucet}.

The first algorithm is the \textit{Conventional Max-min Fairness (CMF)}. This algorithm is implemented as if it operates in the conventional computational setting. The demands are collected for a given block span, which is referred to as an \textit{epoch} in this study. At the beginning of the following epoch these demands are supplied with resources in the MF order by a single node in one step with the \textit{distribute} function.

In the second algorithm, the demands are collected in a given epoch, and the demanders claim their reserved share by calling a \texttt{claim} function in the \textit{claim rounds} of the following epoch. We call this approach \textit{Autonomous Max-min Fairness (AMF)}, since there is no need for a central node to carry out the execution, and the system is operated autonomously by its users. The operation of AMF \textit{emulates} the original algorithm identically, except for the last iteration where the distribution is in the \textit{first come first served} order among overdemanders. Originally, the last iteration is in the \textit{ascending} order of demand volumes, as are all the preceding iterations.

We implemented both the unweighted and the weighted versions of MF for the autonomous case. The reason for not implementing a weighted version of CMF is due to its gas cost structure (demonstrated in Section \ref{cmf_results}). In the following subsections, we give the implementation details of the algorithms explained in this subsection.

\subsection{Conventional Max-min Fairness}
\label{cmf}

As it is in the conventional setting, CMF utilizes two min-heaps, exchanging the demands among each other in each iteration. The operation scheme and the pseudo-code is the same as described in Section \ref{mm_model} (i.e. Figure \ref{max-min_diagram} and Algorithm \ref{mf_pseudo}).

Since Solidity does not offer a built-in data structure for min-heaps, we implemented it during  the development of CMF. We kept the implementation of the min-heap minimalistic in order to keep the gas cost at minimal. Only the amount of demand, and the id of the demanding user is stored and operated on. The remainder of the user attributes are fetched from other data structures when needed (e.g. while writing to the user balance), by using the user id as the key.

We used an array implementation of heap, a complete binary tree, where the values are kept in a node array and the \textit{insert} and \textit{delete minimum} functions are  implemented so that they index and move the nodes according to the min-heap organisation. This is also immune to degeneration attacks, in which case an attacker feeds the tree with selective input to make one branch grow disproportionately, forcing heap functions run in $\mathcal{O}(n)$ instead of $\mathcal{O}(\log n)$ time.

We present the performance of CMF, as well as the min-heap, in Section \ref{cmf_results}.

\subsection{Autonomous Max-min Fairness}
\label{amf}

In AMF, the epochs are divided into \textit{claim rounds}, which are, like the epochs, defined to be a number of successive blocks. At the end of each round, the remaining number of demands, the remaining capacity, and the resulting share is recalculated. The rounds proceed in this manner until either the capacity is depleted, or all demands are supplied. The rounds are used to emulate the iterations of the outer loop (lines $4-20$ of Algorithm \ref{mf_pseudo}) of the distribute function.

In order to avoid repetition, we give the pseudo-code only for the weighted version (WAMF), since it is more general as compared to the unweighted version (AMF), the latter can be seen in Appendix \ref{amf_pseudo}. The pseudo-code of WAMF is presented in Algorithm \ref{wamf_pseudo}. The symbols for the additional variables, and their meanings are given in Table \ref{amf_symbols}. The calculation of weights is obscured from the pseudo-code for the ease of review, and the weights are simply shown as constant variables. The calculation of weights is explained in detail in the next subsection.

In AMF, instead of a single-handedly operating \textit{distribute} function, there is a \texttt{claim} function, which after necessary checks, allows the user assign her allocated share to herself. Each user is expected to execute the function individually, to have carried out the iterations of the inner loop of the \textit{distribute} function (lines $10-18$ of Algorithm \ref{mf_pseudo}), in a decentralised manner.

\begin{figure}
    \centering
    \includegraphics[width=0.9\linewidth]{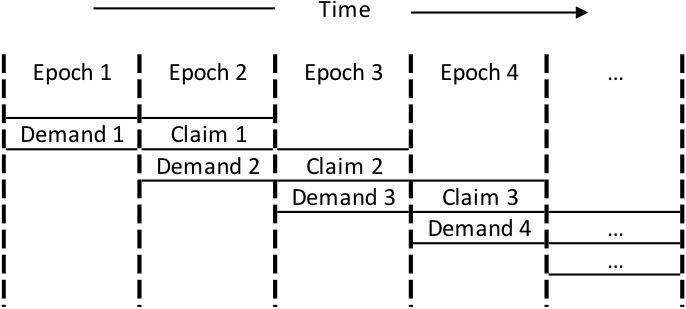}
    \caption{Epochal Layout of Matching Demands and Claims.}
    \label{timeframe}
\end{figure}

Any share unclaimed in its due round/epoch is lost to the user and handed over to the following round/epoch as part of the leftover capacity. In a given epoch, users may make new demands for the next epoch, while claiming their share for the previous. The time frame can be traced in Table \ref{amf_example} over the demands and vertically corresponding claims, and can be seen schematically in Figure \ref{timeframe}.

\begin{table}[ht]
    \caption{An exemplary distribution carried out with AMF}
    \label{amf_example}
    \centering
    \scalebox{0.75}{
        \begin{tabular}{c|c|c|ccc|cc}
            
            \multicolumn{3}{c}{} & User 1    & User 2    & User 3    & Share & Capacity \\
            \hline \hline
            \multirow{4}{*}{} & \multicolumn{2}{c|}{Demand 1} & 4 & 11 & 15 &  & \\
            \cline{2-8}
               & \multirow{3}{*}{}  & Round 1 &  &   &   &   &   \\
            \cline{3-8}
             Epoch 1 & Claim 0 & Round 2 &  &   &   &   &  \\
            \cline{3-8}
                &                           & Round 3 &  &   &   &   &   \\
            \hline
            \hline
            \multirow{4}{*}{} & \multicolumn{2}{c|}{Demand 2} & 11 & 3 & 8 &  & 30\\
            \cline{2-8}
                & \multirow{3}{*}{}  & Round 1 & 4 & 10 & 10  & 10 & 6 \\
            \cline{3-8}
             Epoch 2 & Claim 1 & Round 2 &   &  1 & 3  & 3  & 2 \\
            \cline{3-8}
                &                           & Round 3 &   &    & 2  & 2  & 0 \\
            \hline
            \hline
            \multirow{4}{*}{} & \multicolumn{2}{c|}{Demand 3} & 7  & 8  & 12  & 10 & 30 \\
            \cline{2-8}
                & \multirow{3}{*}{}  & Round 1 & 10 & 3  & 8  & 10 & 9  \\
            \cline{3-8}
             Epoch 3 & Claim 2 & Round 2 & 1  &    &    & 9  & 8  \\
            \cline{3-8}
                &                           & Round 3 &    &    &    &    &    \\
            \hline
            \hline
            \multirow{4}{*}{} & \multicolumn{2}{c|}{Demand 4} & 17 & 13 & 5  &    & 38 \\
            \cline{2-8}
                & \multirow{3}{*}{}  & Round 1 & 7  & 8  & 12 & 12 & 11  \\
            \cline{3-8}
             Epoch 4 & Claim 3 & Round 2 &    &    &    &    &     \\
            \cline{3-8}
                &                           & Round 3 &    &    &    &    &    \\
            \hline
            \hline
           \multirow{4}{*}{} & \multicolumn{2}{c|}{Demand 5}  & .. & .. & .. & .. & 41 \\
            \cline{2-8}
                & \multirow{3}{*}{}  & Round 1 & 13 & 13 & 5 & 13 & 10  \\
            \cline{3-8}
             Epoch 5 & Claim 4 & Round 2 & 4  &    &     & 10  & 6  \\
            \cline{3-8}
                &                           & Round 3 &    &    &    &    &    \\
            \hline
        \end{tabular}
    }
\end{table}

\begin{table}[ht]
	\caption{Symbols used in Algorithm \ref{wamf_pseudo} and their meanings}
    \centering
    \scalebox{0.85}{
        \begin{tabular}{|l|l|} \hline
         Symbol     &   Meaning  \\ \hline \hline
         $EpochCapacity$        &   Amount of replenishment at every epoch, $EpochCapacity \in \mathbb{Z}^+ $  \\ \hline
         $BlockNumber$        &   Current block number  \\ \hline
         $Offset$        &   The block number at which the contract was deployed, offset \\ \hline
         $Epoch$        &   Epoch number  \\ \hline
         $Round$        &   Round number  \\ \hline
         $ResetEpoch$       &   Reset epoch, the epoch at which the total weight was last reset  \\ \hline
         $EpochSpan$       &   Number of blocks in an epoch  \\ \hline
         $RoundSpan$       &   Number of blocks in a round  \\ \hline
         $U$        &   Set of users $U = \{ u_1,\ldots,u_n \}$  \\ \hline
         $TotalWeight[selector]$      &   Total weight for even and odd epochs, $selector \in \{0,1\} $ \\ \hline
         $Volume$        &   Demand volume   \\ \hline
         $User$        &   User  object, $User \in U $ \\ \hline
         $User.demand[selector]$   &   Demand of user $u$ in list $selector$, $selector \in \{0,1\}$  \\ \hline
         $User.demandEpoch[selector]$     &   The last epoch user $u$ made a demand, $selector \in \{0,1\}$ \\ \hline
         $User.claimEpoch$     &   The last epoch user $u$ made a claim \\ \hline
         $User.claimRound$     &   The last round user $u$ made a claim \\ \hline
         $User.balance$      &   Resource balance of user $u$, $User.balance \in \mathbb{Z^+} $  \\ \hline
         $User.weight$      &   Weight of user $u$  \\ \hline
         $Capacity$        &   The existing capacity \\   \hline
\end{tabular}
    }
\label{amf_symbols}
\end{table}

In AMF the demands are kept in a map, rather than a min-heap, since it is necessary for each user to be able to access their own demand entry while claiming it. In the present implementation, the demands are kept for one epoch, and claimed in the following. For this reason, a circular buffer of size two is kept for each user, in order to prevent an incoming demand in a given epoch to overwrite the previous epoch's demand, before it is claimed. This leads to a two dimensional ($n\text{ x }2$) demand vector, where the demands for even and odd epochs are kept separately. Additionally, the variable for keeping the epoch in which the demand was made (for preventing an obsolete demand to interfere with later demands) is implemented; likewise as a circular buffer of size two, in order to separate between the even and the odd epochs.

In addition to the restructured \texttt{demand}, and the newly introduced \texttt{claim} functions, AMF includes a state\footnote{It should be disambiguated that \textit{state} here refers to the state (i.e. values of the global variables at a given time) of the \textit{contract} and not the \textit{blockchain} it runs on.} update function, which is called at the beginning of both. The state update function checks the block number, and calculates the epoch and the round in which the called function will be executed (lines $3$ and $10$, respectively). The number of blocks for the duration of an epoch and a round, is also a parameter of the system, which we experimented on in the present dissertation, and commented on in the results subsection.

The pseudo-code in Algorithm \ref{wamf_pseudo} is organised in three functions, namely, \textit{update state} (lines $1-16$), \texttt{demand} (lines $18-32$), and \texttt{claim} (lines $34-55$). At the beginning of each function (in lines $2$, $20$, and $36$) a local selector variable for the circular buffers is declared and calculated. When called in a given epoch, the state update and the claim functions agree on their selector value, and the demand function assumes its binary complement (e.g. $<0,1,0,1,...>$ for the \textit{state update} and \texttt{claim} functions, and $<1,0,1,0,...>$ for the \texttt{demand} function).

\begin{algorithm}
	\caption{WAMF Pseudocode}
	\begin{algorithmic}[1]
\Procedure{Update State}{$Offset, BlockNumber, Epoch, EpochSpan, RoundSpan$}
\State $selector \gets Epoch ~~mod~~(2)$;
\If{$Epoch < \floor*{ \frac{BlockNumber - Offset}{EpochSpan} }$}
	\State $Epoch \gets \floor*{ \frac{BlockNumber - Offset}{EpochSpan} }$;
	\State $Round \gets \floor*{ \frac{(BlockNumber - Offset) ~~mod~~ (EpochSpan)}{RoundSpan} }$;
	\State $Capacity \gets Capacity + EpochCapacity$\;
	\State $Share \gets \floor*{Capacity / TotalWeight[selector]}$;
	\State return;
\EndIf
\If{$Round < \floor*{ \frac{(BlockNumber - Offset) \% ES}{RS} }$}
	\State $Round \gets \floor*{ \frac{(BlockNumber - Offset) ~~mod~~ (EpochSpan)}{RoundSpan} }$;
	\State $Share \gets Capacity / TotalWeight[selector]$;
	\State return;
\EndIf
\State return;
\EndProcedure
\Procedure{Demand}{$User, Volume$}
\State \Call{updateState}{$Offset, BlockNumber, Epoch, EpochSpan, RoundSpan$}
\State $selector \gets (E + 1)~~mod~~(2)$;
\If{$User.demandEpoch[selector] \neq Epoch$}
	\State $User.demand[selector] \gets Volume$;
	\State $User.demandEpoch[selector] \gets Epoch$;
	\If{$ResetEpoch < Epoch$}
		\State $TotalWeight[selector] \gets User.weight$;
		\State $ResetEpoch \gets Epoch$;
	\Else
		\State $TotalWeight[selector] \gets TotalWeight[selector] + User.weight $;
	\EndIf
\EndIf
\State return;
\EndProcedure
\algstore{wamf}
\end{algorithmic}
\label{wamf_pseudo}
\end{algorithm}

\begin{algorithm}
	\ContinuedFloat
	\caption{WAMF Pseudocode Cont.}
\begin{algorithmic}[1]
\algrestore{wamf}
\Procedure{Claim}{$User$}
\State \Call{updateState}{$Offset, BlockNumber, Epoch, EpochSpan, RoundSpan$}
\State $selector \gets Epoch ~~mod~~(2)$;
\If{$User.demandEpoch[selector] \neq Epoch - 1$ \textbf{or} $Capacity = 0$ \textbf{or} $User.demand[selector] = 0$}
	\State return;
\EndIf
\If{$User.claimEpoch = Epoch$}
	\If{$User.claimRound = Round$}
		\State return;
	\EndIf
\Else
	\State $User.claimEpoch \gets Epoch$;
\EndIf
\State $User.claimRound \gets Round$;
\State $User.balance \gets User.balance + \min{(User.demand[selector], Share * User.weight)}$;
\State $User.demand[selector] \gets User.demand[selector] - \min{(User.demand[selector], Share * User.weight)}$;
\State $Capacity \gets Capacity - \min{(User.demand[selector], Share * User.weight)}$;
\If{$User.demand[selector] = 0$}
	\State $TotalWeight[selector] \gets TotalWeight[selector] - User.weight$;
\EndIf
\State return;
\EndProcedure
\end{algorithmic}

\end{algorithm}

In line $3$, the epoch number is checked for. If the value of $Epoch$ is found to be obsolete, it is updated. Once the epoch number is updated, the round number, the capacity, and the unit share are also updated (lines $5-7$), and the function returns. If epoch number is found to be up-to-date, a similar check is done for the round number in line $10$. This check, when it returns positive, leads to the update of the round number and the unit share (lines $11-12$), and the function returns. If no update is required, the function returns without making any changes in the state.

After updating the state and setting the selector variable, in line $20$ the demand function checks whether the user has made a demand in the then present epoch. If the user has made a demand, the function returns without registering the newly arrived demand. If not, the demand volume ($Volume$) is written to the corresponding slot in the circular demand buffer of the user, and the demand epoch of the user is updated to be the then current epoch (lines $22-23$). In the following line, the function checks whether any demands have been made by other users in the then current epoch. If not, the total weight is set to the user's weight (line $25$), which resets the total weight variable for the next epoch. The variable for keeping the last epoch in which the total weight is reset ($ResetEpoch$) is updated in line $26$. If demands have been made by other users prior to the then current call (i.e. $ResetEpoch = Epoch$) the weight of the user is added to the total weight, to be accounted for in the next epoch (line $28$).

The claim function, similar to the demand function, starts with updating the state and initiating the selector variable. It continues with a number of checks. Unless the demand has been done in the previous epoch and is greater than $0$, or if the capacity is depleted, the function returns without taking any further action. Following that in line $40$ the function checks whether the user has made any claims in the then current epoch. If so, the last round the user made a claim is checked (line $41$). If that also turns positive, which means the user has claimed her fair share for the round, the function returns without making any assignments.

If the check in line $40$ turns out negative, meaning this is the user's first claim in the then present epoch, the variable for the last epoch the user made a claim is updated (line $45$). After that, a similar variable for the round is updated in line $47$. Next, the assignment operations similar to the ones in Algorithm \ref{mf_pseudo} is done in lines $48-50$.

Note that this algorithm differs from the CMF algorithm in that the leftover demands are not inserted into another heap; they remain in the map. Instead, the fully satisfied demands are removed from the cumulative weight variable in lines $51-53$, having the same effect as deleting the minimum in CMF algorithm. This way, as long as there is an unsatisfied demand, the user's weight is included in the total weight, and the unit share is calculated accordingly. At the end of the epoch, all demands are obsoleted.

\subsection{Weighted Autonomous Max-min Fairness}
\label{wamf}

As the operation of the algorithm is described in Section~\ref{amf}, the only part that is left to be explained in this subsubsection is the calculation of weights.

We defined weights to be the multiplicative inverses of the total demand volume, up to and including the then present demand. The reason for our choice of this weighting policy is to incentivise the users to make the minimum demands that can satisfy their needs. It is achieved due to the fact that in this setting the most rational behavior of the user is to keep her demand minimal, in order not to be disadvantageous in the long run.

For comparison, an alternative policy would be to weight the users inversely proportional to the total volume of previously allocated resources, which would lead the distribution of the total allocated volume of the resources among the users to tend to a uniform distribution in the long run. This is a matter of the needs of the system that the algorithm will be adopted to serve to.

In order to weight the users inversely proportional to the total demand volume up to and including their by then present demand ($dt_u$), the multiplicative reciprocal of $dt_u$ is calculated. This poses a problem in the smart contract context, since Solidity does not offer floating point data types. In other words, since the demand volumes are defined to be positive integers, it is not possible to keep weights as they are, since the value needs floating point data type to be stored. Instead, we keep the total demand volume for each user ($dt_u$ for user $u$), introduce an intermediary variable $p$ (standing for \textit{precision}) and take the weight equal to:

\[w_u = \floor*{\frac{p}{dt_u}}\]

We get rid of this intermediary variable while calculating the unit share. Therefore, instead of

\[s = \floor*{\frac{c}{\sum_{u=1}^{n} w_u \cdot I(d_u)}}\]
\noindent 
we use:

\[s = \floor*{\frac{c \cdot p}{\sum_{u=1}^{n} w_u \cdot I(d_u)}}\]
\noindent 
since

\[s = \floor*{\frac{c \cdot p}{\sum_{u=1}^{n} \frac{p}{dt_u} \cdot I(d_u)}} = \floor*{\frac{c}{\sum_{u=1}^{n} \frac{1}{dt_u} \cdot I(d_u)}} \]
\noindent where $I(x)$ is the indicator function, which returns $1$ if $x$ is a positive number, and $0$ if $x$ equals $0$. In this context it allows us to indicate that only the weights of users who made a demand are included in the total sum. Similarly, while calculating the user share we use the intermediary variable $p$:

\[s_u = \floor*{\frac{s \cdot \floor*{\frac{p}{dt_u}}}{p}}\]
\noindent 
As long as the value of $p$ is larger than the total demand volume of the user, we obtain non-zero weights from $\floor*{\frac{p}{dt_u}}$. For $p = 10^k, k \in \mathcal{Z^+}$ is the number of decimal places stored for weights.

\section{Procedure and Parameters}
\label{procedure_and_parameters}

We implemented our algorithms in Solidity programming language and run on an EVM environment \cite{dannen2017introducing}, and more specifically, its Parity implementation, as mentioned above. The main reason for selecting this framework is its wide use among blockchain ecosystems. Many blockchain ecosystems and blockchain based systems utilise either EVM or virtual machines similar to EVM, and support Solidity programming language for smart contracts (e.g. \cite{baird2018hedera, tron, cheng2019ekiden, niloy2021blockchain}), and for this reason there are also studies available on the performance \cite{wohrer2018design}, security \cite{wohrer2018smart}, and inspection \cite{bragagnolo2018smartinspect}. It is a high-level, easy to read, object oriented script language.

We tested our algorithms in a local blockchain, operated by Parity Ethereum $2.7.2$ \cite{parity}, and we implemented the smart contracts in Solidity $0.5.13$. The block gas limit we assume is $8.000.000$ units\footnote{This value is taken from the Ethereum's actual block gas limit at the time we started our experiments. The value is dynamically set and updated in each system by the collective decision of its miners, thus it is not up-to-date. Nevertheless, for the purpose of the present study, it does not constitute a problem, and even an up-to-date value will be obsoleted in the tme of the readers view so we left it as it is.}.

Parity implementation of Ethereum offers customisable consensus protocols. Among those is the so-called \textit{instant seal engine}, which places each transaction into an individual block of its own. The engine is specifically designed for contract development, since the block latency is rarely a relevant parameter in the development and verification processes of the algorithms, at least for the time being.

For our case, the instant seal engine also allows us operationalise \textit{time} in terms of number of blocks and, in turn, define the \textit{epoch span} and the \textit{round span} in terms of it. As such, our results are generalisable to every blockchain environment, independent of the consensus algorithms and parameters they employ.

\subsection{Timing and Synchronisation}
\label{synchronisation}

In a setting with $n$ users, in the first epoch, $n$ blocks are used for user registration function calls and $2n$ blocks are filled with empty transactions in order to synchronise the process. The following demand function calls occupied $n$ more blocks, concluding the first epoch. From the second epoch on, the sequence is $3$ rounds of claim in $3n$ blocks, followed by $n$ blocks of demand for the next epoch. Therefore the span of a round is chosen to be equal to $n$ blocks, and an epoch equal to $4n$ blocks. The tests are run for $3$ sets, each extended over $4$ epochs as described above. Averages of each set are collected, and averaged out for the final result to be reported. The parameters may also be reviewed in Table \ref{amf_parameters}.

\begin{table}[ht]
    \caption{The values used in the tests for AMF and WAMF.}
    \label{amf_parameters}
    \centering
    \begin{tabular}{rrp{3.5cm}}
    Parameter       & Value & Definition \\
    \hline
    \hline
    Number of Users & $n$     & The number of users in the system\\
    \hline
    Epoch Capacity  & $20n$  & The amount to be distributed for each epoch\\
    \hline
    Epoch Span      & $4n$   & The duration of an epoch in number of blocks\\
    \hline
    Round Span      & $n$     & The duration of a round in number of blocks\\
    \hline
    Demand Interval & $[10,30)$ & The interval from which the demands are drawn\\
    \hline
    \end{tabular}
\end{table}

\section{Results}
\label{results_1}

The results of the tests carried out for CMF and W/AMF are presented in the following subsections. The data are available in \cite{metin2020faucet}.

\subsection{CMF Results}
\label{cmf_results}

As indicated in Section \ref{cmf}, in the CMF, the demand vector is implemented as an array of two min-heaps, exchanging the demands among each other at each iteration. The demands arriving from the users are collected in $D_0$ for the span of an epoch. At the end of the epoch, the distribute function is called by the authority node, and the distribution is done. The first iteration is done over $D_0$, taking all demands from the smallest to the largest, granting the available share to the user, and finally either deleting the minimum demand, if it is completely supplied, or deleting it from $D_0$ and inserting it to $D_1$, otherwise, to be supplied in the next iterations if possible. The heaps exchange functions, and the process is repeated until either all the demands are supplied, or the capacity for the epoch is exhausted (see Algorithm \ref{mf_pseudo}).

Gas usage averages for $n=100$ entry sets are shown in Table \ref{heap_results}. For comparison, the gas performance of a general case heap implementation~\cite{zmitton}, called Eth-heap,  is provided next to our results:

\begin{table}[ht]
    \caption{Average gas costs for \textit{Insert} and \textit{Delete Minimum} functions}
    \label{heap_results}
    \centering
    \begin{tabular}{rrr}
    Function 		& Eth-heap	& Present Study	\\
    \hline
    \hline
    Insert			& 101.261	&  95.459		\\
    \hline
    Delete Minimum	& 170.448	& 133.272		\\
    \hline
    \end{tabular}
\end{table}

Considering the $8.000.000$ block gas limit, the heap operations impose an upper bound of $60$ entries to be processed per block, on average, as seen with the cost of operations in Table \ref{heap_results}. This number is to be further lowered  with the additional cost of assignment operations, needed to record the fair share of each user to her balance.

The finding immediately implies that an algorithm implemented as a smart contract and relying on a central node to carry out the distribute function, cannot support more than $\sim10$ users, assuming that $3$ iterations are necessary on average for a distribution process to complete. The exact number is a function of how disperse the demands are, since the number of delete/insert operations is dependent on the number of iterations necessary to answer all the demands, which in turn is dependent on how disperse the demands are.

This is also the reason why a weighted version of CMF has not been implemented in the present dissertation. The extra cost of calculating and storing weights would make the weighted version perform even worse than the unweighted version.

\subsection{AMF and WAMF Results}
\label{amf_results}

The first advantage to be pointed out for AMF is that it virtually has no limit for the number of users that the system can support. The average gas costs of \texttt{demand} and \texttt{claim} functions for a system with $10, 50, 100$ and $500$ users can be seen in Table \ref{amf_performance}. The tests have been carried over in a setting where users have made demands, and claimed their demands in the succeeding epoch. The results indicate that several \texttt{demand} and \texttt{claim} function calls can be included within a block, without running into the block gas limit exhaustion problem.

\clearpage

\begin{table}[ht]
    \caption{Average and total gas costs of W/AMF \textit{demand} and \textit{claim} functions.}
    \centering
    \scalebox{0.91}{
        \begin{tabular}{c|c|cc}
        Function    & No. of Users    & AMF & WAMF \\ 
        \hline  \hline
        \multirow{4}{*}{Demand} & $10$  & $70.245$ & $79.732$ \\ \cline{2-4}
                                & $50$  & $67.351$ & $77.135$ \\ \cline{2-4}
                                & $100$ & $66.989$ & $76.835$ \\ \cline{2-4}
                                & $500$ & $66.700$ & $71.365$ \\ \hline \hline
        \multirow{4}{*}{}       & $10$  & $46.800/140.401$ & $46.643/145.931$\\ \cline{2-4}
                    Claim       & $50$  & $42.240/126.720$ & $44.852/134.558$\\ \cline{2-4}
                (Avg./Total)    & $100$ & $42.114/126.344$ & $44.763/134.289$\\ \cline{2-4}
                                & $500$ & $42.047/126.143$ & $45.319/135.959$\\ \hline
        \end{tabular}
    }
    \label{amf_performance}
\end{table}

The results also indicate that the cost of \texttt{demand} and \texttt{claim} functions do not grow with the growing number of users. On the contrary, there is a slight decrease in the average costs, with the growing number of users. The reason for this is the fact that in each epoch the first call to both functions are costlier, since state variables are updated in these calls. With large sample sizes, this difference tends to even out better as compared to the relatively smaller sample sizes.

One thing that should be accounted for is that the average cost of demand function declines throughout the rounds. The reason for this is, some demands have been fully supplied in the previous epoch, thus, fewer calls to claim function lead to the full execution of the function (i.e. calls from users whose demands have already been satisfied return without making any assignments). The average claim costs of rounds for Max-min and Weighted MF schemes can be seen in Table \ref{claim}.

\begin{table}[ht]
    \caption{The cost of the \textit{claim} function over rounds ($n=500$).}
    \label{claim}
    \centering
    \begin{tabular}{rrr}
    \multicolumn{1}{c}{Round}   & \multicolumn{1}{c}{AMF}   & \multicolumn{1}{c}{WAMF}\\
    \hline
    \hline
    $1$ & $64.677$ & $67.211$\\
    \hline
    $2$ & $32.717$  & $36.158$\\
    \hline
    $3$ & $28.749$  & $32.589$\\
    \hline
    Average & $42.047$  & $45.319$\\
    \hline
    Total & $126.143$  & $135.959$     \\
    \hline
    \end{tabular}
\end{table}

The number of rounds, as indicated in Section \ref{cmf_results} for the number of iterations of CMF, is a function of the initial distribution of the demands. In our tests, we drew random demands from an approximately uniform distribution offered by Javascript Math.random() function, in the range $[10,30)$, and the epoch capacity is set to $20n$, so that on average the overdemands and underdemands could balance each other out.

In all the simulations the distribution is completed in $3$ iterations. Therefore, in the tests presented here, we run the system for $3$ rounds of claims. The results are cross-checked with the Python simulations and proved identical. We suspect that with the parameters used in this study, $3$ iterations might be an upper bound, but we do not have a proof. Further investigation needs to be carried out to in order to come up with a theoretical
bound.

Another variable that can be parameterised according to the policy and that would effect gas costs is the size of the variables used to represent amounts. The size of the variables can be chosen smaller to save from the extra cost of unused space. The necessary sizes for the variables is dependent on the total amount that is planned to be distributed in the long run, maximum available allocation in an epoch, the maximum number of epochs to distribute all the resource. In the present dissertation, all the variables are implemented as their $256$ bit defaults, in order not to lose generality.

\chapter{MAX-MIN FAIRNESS RESTRUCTURED}
\label{max-min_restructured}

In this Chapter we will present four algorithms that we obtain by restructuring the MF algorithm. The operation of these algorithms are different from the original MF algorithm, but for any given input, the output they produce is identical to the output MF produces.

\section{Present Models}
\label{present_models}

In this subsection, we will describe the working and the domain of QMF and SMF models, and also their weighted counterparts, in comparison with the conventional MF model. In contrast to the conventional distribution algorithm, both algorithms presented in the present dissertation calculate and declare the maximum share that the system has to offer, so that when the users are allowed to take the minimum of this declared share and their demands (i.e. $\min\{s,d_u\}$), the capacity at hand shall not be exceeded. In turn the share is declared, the users are allowed (and expected) to assign this minimum to their balances, individually.

In the weighted versions, the same procedure is carried out for calculating a \textit{unit share}, with which each user can obtain their individually proposed \textit{user share} by multiplying it with their individually assigned \textit{user weight}s (i.e. $s_u = s * w_u$); and then they can take the minimum of their user share and their demand (i.e. $\min\{s_u, d_u\}$).

What differentiates QMF and SMF is the procedure each one utilizes to calculate the share, which we discuss in detail in the subsequent subsections. Before moving on to describe the particular details of the two models, we will continue with their common constructs.


Both models operate in the same temporal setting. As in the MF models presented before, in these settings also, the time is fractured into \textit{epochs}, which is defined to be a collection of a fixed number of successive blocks. For the duration of an epoch, users are allowed to make demands. At the end of an epoch, the available share is calculated according to the accumulated demands, and in the following epoch, the users can claim their fair share during the span of the epoch. The users are also allowed to make new demands to be collected in the following epoch, while claiming their demands submitted in the preceding epoch. The time overlay of demands and claims may be seen clearer in Figure \ref{timeframe}. The right to any share unclaimed in its due epoch is lost, and the unclaimed share is added to the capacity of the next epoch, along with the leftover capacity, if there are any.

The state of the system is accounted for, again, by a dedicated function, \textit{update state}, which is called at the beginning of both the \texttt{demand} and the \texttt{claim} functions. \textit{update state} checks the validity of the epoch number with respect to the block number. If the epoch number needs to be updated, the capacity is replenished according to a predefined policy, and the share is recalculated. For simplicity we kept this policy in its simplest and replenished the capacity by a \textit{constant} amount, which we refer here to as the \textit{Epoch Capacity}. The function \textit{update state} does not explicitly invalidate the obsoleted demands; they are rather invalidated by the update of the \textit{epoch} variable, by the virtue of the organisation of the remaining functions and the data structures that represent the demands.

In order to recalculate the share, \textit{update state} accesses a view function, \textit{calculate share}. The main difference between QMF and SMF, and their weighted counterparts also, is the structure of their respective \textit{calculate share} functions, which will be explained in detail in the relevant subsections following.

As indicated above, in both implementations \textit{calculate share} is declared to be a \textit{view function}, which means that the function does not store any data on the permanent storage variables of the contract. It should be noted that storing data on the permanent storage variables, which is colloquially called a \textit{storage write}, is the costliest operation\footnote{For comparison, the second costliest operation is \textit{storage read}, and there is more than an order of magnitude between the cost of the two: 20.000 vs. 800 gas per operation.} in terms of gas expenditure. Considering the fact that the cost of the \textit{calculate share} function is the main bottleneck for remaining within the boundaries of block gas limit, avoiding storage write operations is crucial.

\clearpage

Both implementations rely on iterating over the user demands, and both get their numeric limitations over the efficiency of the loops for these iterations. The abstractions for and the layout of the demand data, in turn, determine the efficiency of these loops. In W/QMF, the loop iterates over the number of demands for the predefined demand volume interval (i.e. $[1,Quanta]$), and in W/SMF over the user demands (i.e. $\{d_1,...,d_n\}$), hence the limitations on demand volume and the number of users, respectively.

\subsection{Quantized Max-min Fairness Model}
\label{qmf_model}

The operation of QMF is analogous to that of \textit{Counting Sort} algorithm, in which case to sort a collection of elements in a predefined interval, the algorithm traces the number of occurrences of each element, and enumerates the sorted list according to those counts. Likewise, QMF traces the number of demands for each demand volume, in a predefined demand volume interval, and calculates the share over these counts.

When the \textit{calculate share} function is called, the number of demands for each demand volume is ready, since this part is handled by the \texttt{demand} function. When a demand arrives, the number of demands for the relevant demand volume is incremented by $1$, in addition to the other operations for recording the demand (e.g. updating the \texttt{demand} variable in the user list). Conveniently, demands are represented with an array, instead of a heap, since random access to demand volume counts are needed to record the increments.

The main loop of the \textit{calculate share} iterates over the demand array, starts by proposing $1$ as the share and calculates the total capacity needed to declare the share as such. If the capacity is sufficient, the next iteration is taken, until reaching a proposal which would lead to a shortage of capacity. The loop breaks when it reaches such a proposal, and the penultimate proposal is returned to the calling function (i.e. \textit{update state}) to be declared as the share.

\clearpage

The formula for calculating the total necessary capacity for a proposal $p$ ($1 \leq p \leq q,\; p, q \in \mathbb{Z}$) is:

\begin{equation}\label{qmf_loop}
    \begin{split}
        &\sum_{i=1}^{p-1} i \cdot d_i + \sum_{j=p}^q p \cdot d_j\\
        =&\sum_{i=1}^{p-1} i \cdot d_i + p \cdot \sum_{j=p}^q d_j\\
        =&\sum_{i=1}^{p-1} i \cdot d_i + p \cdot (D - \sum_{i=1}^{p-1} d_i)
    \end{split}
\end{equation}

\noindent where $d_i$ stands for the $i$-th entry in the demand array, and $D$ for the total number of demands, which is collected during the demand epoch by the \texttt{demand} function. The remaining terms are calculated within the loop, each new iteration using the previous iteration's cumulative values. The first term of the equation stands for the capacity reserved for the underdemanders, and the second term for the capacity available for the overdemanders. A numerical example can be seen in Table \ref{qmf_example}.

\begin{table}[ht]
    \centering
    \caption{QMF Procession Example ($c = 50$)}
        \begin{tabular}{rrrrrrrr}
    \toprule
    Demand Volume      & 1  & 2  & 3  & 4  & 5  & 6  & 7  \\ \midrule
    Number of Demands (NoD)   & 3  & 2  & 1  & 0  & 3  & 0  & 4  \\
    NoD Cumulative   & 3  & 5  & 6  & 6  & 9  & 9  & 13 \\
    Total Demand Volume (TDV)      & 3  & 4  & 3  & 0  & 15 & 0  & 28 \\
    TDV Cumulative     & 3  & 7  & 10 & 10 & 25 & 25 & 53 \\
    Necessary Capacity  & 13 & 23 & 31 & 38 & 45 & \boxed{49} & 53 \\
    \bottomrule
    \end{tabular}
    \label{qmf_example}
\end{table}

\subsection{Weighted Quantized Max-min Fairness Model}
\label{wqmf_model}

The main difference in WQMF is that instead of a globally defined \textit{share} for all users, we calculate \textit{unit share} and each user's individual share is calculated by multiplying the unit share with the user's individual weight. Unit share is defined as the share reserved for unit weight (i.e. $w = 1$). According to this:

\[
s_u = w_u \cdot \frac{c}{\sum_{u = 1}^{n} w_u \cdot I(d_u)}
\]

\noindent where $s_u$ denotes the share and $w_u$ denotes the weight of the user $u$, and $n$ is the total number of users in the system. $I(x)$ is the indicator function, which returns $1$ if $x$ is a positive number, and $0$ if $x$ equals $0$. In this context it allows us to indicate that only the weights of users who made a demand are included in the total sum.

In order to calculate maximum available unit share, in the \texttt{demand} function we calculate and keep the minimum unit share that suffices to satisfy the user's demand. This is given by:

\[i = \ceil*{\frac{d_u}{w_u}}\]

\noindent where, $i$ stands for the \textit{index} to be updated in the demand array. In addition to the demand array, we also utilize a weight array, and $i$-th entry in both arrays are incremented by their corresponding values (i.e. by $d_u$ and by $w_u$, respectively), instead of by $1$, since total demand volume and total weight values are needed in the calculation, instead of the total number of demands. Similar to QMF, the necessary capacity for declaring the unit share as $p$ ($1 \leq p \leq q,\; p, q \in \mathbb{Z}$) is then given by:

\[\sum_{i = 1}^{p - 1} d_i + p \cdot \sum_{i = p}^{q} w_i\]

As it is in QMF, the first term gives the total supply volume satisfying the underdemanders, and the second term gives the capacity available for the overdemanders, if the unit share is to be declared as $p$.

In order to iteratively calculate the necessary capacity for all $i$ and select the maximum available value, we manipulate the second term and calculate:

\[\sum_{i = 1}^{p - 1} d_i + p \cdot (W - \sum_{i = 1}^{p - 1} w_i)\]

\noindent where $W$, analogous to $TD$ in QMF, is the total weight of the users that made a demand in the previous epoch, which we collect and calculate during the demand epoch within the \texttt{demand} function.

It should be noted that the size of the demand and weight arrays should be equal to the range of possible index values. Since the range of the available demand volumes are restricted in the $[1,q]$ interval, the algorithm needs the range of the available weight values also be a finite set to be able to operate. This brings about the further restriction that the image of the weighting function be a finite interval.

\subsection{Simulated Max-min Fairness Model}
\label{smf_model}

The operation of the \textit{calculate share} function of SMF is almost identical to that of the conventional MF algorithm. The mere difference is that the iterative assignments are replaced with a single update to a \textit{memory variable}, which is significantly more affordable in terms of gas expenditure as compared to its \textit{storage} counterpart, in order to calculate the maximum available share that the system has to offer to each user without exceeding the capacity. The users, then, assign the minimum of their demands and the share, individually.

\clearpage

The reason for the decoupling of calculating the share and assigning it to the user balances is the cost of storage write operation, as explained in Section \ref{present_models}. Although the share of each user is calculated during the operation, it is a better strategy in terms of gas cost, to not keep this information, and handle the individual assignments in a separate \texttt{claim} function. In fact, as it is shown in Section \ref{cmf}, the alternative leads to rapid block gas limit exhaustion, and the system cannot support more than a few users.

SMF iterates over the user demand vector and checks the demand of each user individually, collecting all the valid demands in a \textit{memory heap}. This heap is a binary complete tree, implemented as an integer array on which two functions operate, one for inserting new values and the other for removing the minimum element, which always reside in the tree root. These are what is called \textit{pure} functions in the Solidity Programming Language, which do not perform neither storage write nor storage read operations, and as such, they are expected to be the least costly family of operations. The array is removed from the memory upon the return of the \textit{calculate share} function.

\begin{figure}
    \centering
    \begin{tikzpicture}
    \node [draw,rectangle,align=center] (dho) {Demand\\Heap $0$};
    \node [draw,rectangle,right = of dho,align=center] (dhi) {Demand\\Heap $1$};
    \node [draw,rectangle,align=center] (shr) at ($(dho)!.5!(dhi)+(0,-3)$) {Final Share};
    
    \draw [->,dashed] ($(dho.north)!.5!(dho.north east)$) to [bend left, above, out=60, in=120] node [align=center] {Leftover\\Demand} ($(dhi.north west)!.5!(dhi.north)$);
    \draw [<-,dashed] ($(dho.south)!.5!(dho.south east)$) to [bend left, below, out=-60, in=-120] ($(dhi.south west)!.5!(dhi.south)$);
    
    \draw [->] (dho) to [bend right] node [left,align=left] {+} node [right,align=right] {Partial} (shr);
    \draw [->] (dhi) to [bend left] node [left,align=left] {Shares} node [right,align=right] {+} (shr);
    
    \draw[thick] ($(current bounding box.south west) - (.5,.5)$) rectangle ($(current bounding box.north east) + (.5,.5)$);
    
    \draw [<-,thick] (dho.west) --++ (-1,0) node [left,align=center] {User\\Demands};
    \draw [->,thick] ($(current bounding box.east)!.5!(current bounding box.south east)$) --++ (1.5,0) node [midway,above,align=center] {Leftover\\Capacity};
\end{tikzpicture}
    \caption{SMF Operation Diagram}
    \label{smf_diagram}
\end{figure}
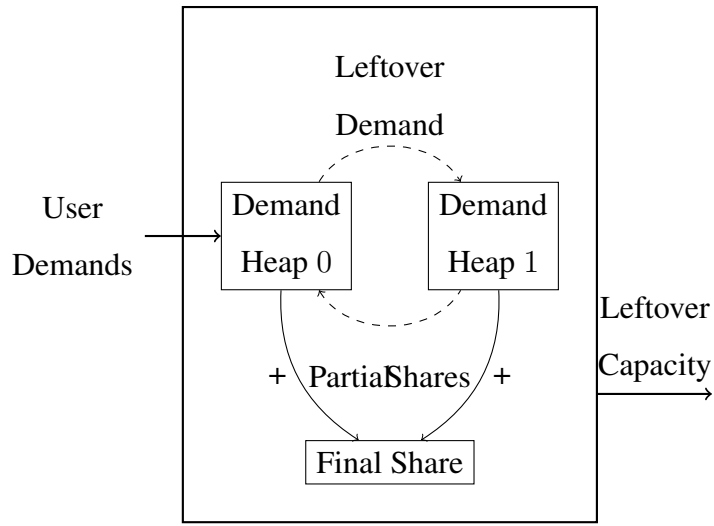

The \textit{calculate share} function of SMF inserts only the demand volume to the minimum heap $D_0$. This is because the owner of the demand is not needed, since the assignment operation will not be handled here. Once $D_0$ is populated, the remainder of the functioning is identical to MF, as indicated before, with the exception of the assignment operations. The procedure is represented in Figure \ref{smf_diagram} visually. In the figure, \textit{partial shares} refer to the share at each iteration, which is added to the \textit{final share}\footnote{In Algorithm \ref{smf_pseudo}, this is represented with the \textit{result} variable}, the variable to be updated and returned to the calling function of \textit{calculate share}.

\subsection{Weighted Simulated Max-min Fairness Model}

In contrast with SMF, WSMF utilizes a minimum heap of a node struct, rather than a simple heap of integers, to represent the demand volumes. This node struct keeps the weight of the user, in addition to the demand volumes, since the total weight is needed in the calculation of \textit{unit share}, as explained in Section \ref{mm_model}. In agreement with SMF, WSMF does not keep user id variable in the calculation loop. An additional difference with SMF is that, the unit share is multiplied with the user weight within the \texttt{claim} function. Other than these differences, the operation of the two algorithms are identical.

\subsection{Weighting Policy}
\label{weighting}

In the present dissertation, we implemented and experimented W/SMF with two different weighting policies as it will be seen in Section \ref{results_2}. In the first case, we chose the weights constant, randomly drawn for each user in a predefined weight interval. In the second case, we dynamically weighted each user, inversely proportional to their cumulative demand volumes, up to and including the then present demand. The first is the trivial case and it is implemented as a basis for the comparison of the added cost of calculating the dynamic weights of the second case.

\section{Implementation}
\label{qs_implementation}

In the following subsections, we will explain the implementations of QMF and SMF in detail, over the pseudocodes created for each. The reason for choosing the unweighted versions to be explained in detail is brevity. The reader might access weighted pseudocodes in Appendices \ref{wqmf_pseudo} and \ref{wsmf_pseudo}, which we believe will be readily intelligible once the unweighted code is examined.

We also note that the actual smart contracts, which can be accessed in the repository at \cite{metin2020faucet}, includes additional functions to the ones explained in the following subsections, for registering users, withdrawing currency etc., which the distribution process operates independent of. They have been implemented for convenience, and to demonstrate how the system can operate with a simple interface, and as such their performance is not a relevant metric for the overall operation of the system. Therefore, they are not included in the pseudocode, and not explained in the text.

\subsection{Quantized Max-min Fairness}
\label{qmf}

QMF (Algorithm \ref{qmf_pseudo}) consists of four functions, two of which the user has access to, and the other two are accessed within the former.

\begin{algorithm}
	\caption{QMF Pseudocode}
	\begin{algorithmic}[1]
\Procedure{Demand}{User, Volume}\Comment{Make a Demand}
\State \Call{Update\_State()}{};
\State $selector \gets (Epoch + 1) \pmod{2}$;
\If{$User.demandEpoch[selector] = Epoch$}
    \State return;
\EndIf
\State $User.demandEpoch[selector] \gets Epoch$;
\If{$ResetEpoch[selector][Volume] \neq Epoch$}
    \State $ResetEpoch[selector][Volume] \gets Epoch$;
    \State $Demands[selector][Volume] \gets 1$;
\Else
    \State $Demands[selector][Volume]++$;
\EndIf
\State $User.demand[selector] \gets Volume$;
\State $TotalDemands++$;
\EndProcedure
\Procedure{Claim}{User}\Comment{Claim User Share}
\State \Call{Update\_State()}{};
\State $selector \gets Epoch \pmod{2}$;
\If{$User.demandEpoch[selector] \neq Epoch - 1$}
    \State return;
\EndIf
\If{$User.claimEpoch = Epoch$}
    \State return;
\EndIf
\State $User.claimEpoch \gets Epoch$;
\State $User.balance \gets min(Share, User.demand[selector])$;
\State $Capacity \gets Capacity - min(Share, User.demand[selector])$;
\EndProcedure
\algstore{qmf}
\end{algorithmic}
	\label{qmf_pseudo}
\end{algorithm}

\begin{algorithm}
	\ContinuedFloat
	\caption{QMF Pseudocode Cont.}
	\begin{algorithmic}[1]
\algrestore{qmf}
\Procedure{Update\_State()}{}
\If{$Epoch \neq \frac{BlockNumber - Offset}{EpochSpan}$}
\State $Epoch \gets \frac{BlockNumber - Offset}{EpochSpan}$;
\State $Capacity \gets Capacity + EpochCapacity$;
\State $Share \gets \Call{Calculate\_Share()}{}$;
\State $TotalDemands[(Epoch + 1) \pmod{2}] \gets 0$;
\EndIf
\EndProcedure
\Procedure{Calculate\_Share()}{}
\State $selector \gets Epoch \pmod{2}$;
\State $cumulativeDemands \gets 0$;
\State $cumulativeDemandVolume \gets 0$;
\For{$i \gets 1, Quanta$}
    \If{$ResetEpoch[selector][i] = Epoch - 1$}
        \State $cumulativeDemands \gets cumulativeDemands + Demands[selector][i]$;
        \State $cumulativeDemandVolume \gets cumulativeDemandVolume + $
        \item[] \hfill $i * Demands[selector][i]$;
    \EndIf
    \If{$Capacity < cumulativeDemandVolume + i * (TotalDemands[selector] - $ \hfill
    \item[] \hfill $cumulativeDemands) $}
        \State \Return $i - 1$;
    \EndIf
\EndFor
\State \Return $Quanta$;
\EndProcedure
\end{algorithmic}
\end{algorithm}

The \texttt{demand} function takes the user id and demand volume as arguments, and starts by updating the state by a call to \textit{update state} function. In order to select the right portion of the circular buffers, a selector variable is initiated once the state is updated. In the circular buffers, the \texttt{demand} function writes to $D_0$ in odd epochs, and to $D_1$, in even epochs (line $3$).

The function then proceeds to check whether the user has already made a demand in the then present epoch. If so, it returns without taking any further action, and if not, proceeds to record the demand. The variable for keeping the epoch at which the user made the last demand is updated to be the then current epoch (line $7$).

In lines $8-13$, the function checks whether or not the relevant entry in the demand vector has been updated in the then current epoch. If it is the first time the entry will be updated in the then present epoch, it is set to $1$, and the relevant entry in the \textit{demand reset array} (the array for keeping at which epoch the relevant entry in the demand array has been reset) is updated to be the then present epoch. If the entry is found out to be updated before, it is incremented by $1$. Following that the demand volume is recorded in the user demand vector and the number of total demands is incremented by $1$ (lines $14-15$); then the function returns.

The \texttt{claim} function starts with checks and updates on the epoch variable (lines $19-25$), similar to the ones in the \texttt{demand} function. In claim function, however, $D_0$ is used in the even epochs, and $D_1$ in odd ones. This alternating pattern enables \texttt{demand} and \texttt{claim} functions run in the same epochs, without interfering in each others operation. The \textit{update state} and the \textit{calculate share} functions agree with the \texttt{claim} function in the parity of their selector variables. The function continues with updating the user claim epoch. Finally, it returns after assigning the minimum of the share and the user demand to the user account (line $27$), and discounting that amount from the capacity (line $28$).

The \textit{update state} is an \textit{internal} function, as seen before, called by \texttt{demand} and \texttt{claim} functions. It is mainly responsible for updating the epoch (lines $31-32$), and if the epoch needs to be updated, \textit{capacity} (line $33$), \textit{share} (line $34$), and \textit{totalDemands} (line $35$) variables along with it. The due epoch number is calculated by subtracting the number of the block that the contract was deployed (\textit{offset}) from the then current block number, dividing it by the epoch span, and finally taking the floor of the resulting number (line $31$).

The function \textit{calculate share} is accessed only within the \textit{update state} function, thus it assumes the state to be up-to-date, and immediately starts with initiating the selector variable, which, as mentioned before, agrees with the selector variable of the claim function.

Having initiated the selector variable, the function initiates two more local variables. These variables are used to keep the cumulative number of demands and the cumulative demand volume, as the share is calculated iteratively, thus the names of the variables: \textit{cumulativeDemands} and \textit{cumulativeDemandVolume}.

Lines $42-50$ show the main loop of the \textit{calculate share} function, which at each iteration, calculates the cost of declaring the share as equal to the number of the iteration. That is to say, in first iteration the cost of declaring the share as $1$ is calculated, in second iteration $2$, and so on, up to the maximum allowed demand volume, \textit{Quanta}. If at any step the cost exceeds the available capacity, the loop breaks, returning the penultimate proposal as the share. If the loop finishes without breaking, the value \textit{Quanta} is returned.

Since the function keeps the cumulative values in the local variables, the write function is not costly. The main cost is due to the storage reads in lines $44$ and $50$, which is still affordable, as shown in Section \ref{results_2}.

\subsection{Simulated Max-min Fairness}
\label{smf}

Like QMF, SMF (Algorithm \ref{smf_pseudo}) also consists of four functions, two of which the user has access to, and the remaining two is accessed within the former.

\begin{algorithm}
	\caption{SMF Pseudocode}
	\label{smf_pseudo}
	\begin{algorithmic}[1]
\Procedure{Demand}{User, Volume}\Comment{Make a Demand}
\State \Call{Update\_State()}{};
\State $selector \gets (epoch + 1) \pmod{2}$;
\If{$User.demandEpoch[selector] = Epoch$}
    \State return;
\EndIf
\State $User.demand[selector] \gets Volume$;
\State $User.demandEpoch[selector] \gets Epoch$;
\EndProcedure
\Procedure{Claim}{User}\Comment{Claim User Share}
\State \Call{Update\_State()}{};
\State $selector \gets epoch \pmod{2}$;
\If{$User.demandEpoch[selector] \neq Epoch - 1$}
    \State return;
\EndIf
\If{$User.claimEpoch = Epoch$}
    \State return;
\EndIf
\State $User.claimEpoch \gets Epoch$;
\State $User.balance \gets min(Share, User.demand[selector])$;
\State $Capacity \gets Capacity - min(Share, User.demand[selector])$;
\EndProcedure
\Procedure{Update\_State()}{}
\If{$Epoch \neq \frac{BlockNumber - Offset}{EpochSpan}$}
\State $Epoch \gets \frac{BlockNumber - Offset}{EpochSpan}$;
\State $Capacity \gets Capacity + EpochCapacity$;
\State $Share \gets \Call{Calculate\_Share()}{}$;
\EndIf
\EndProcedure
\algstore{smf}
\end{algorithmic}
\end{algorithm}

\begin{algorithm}
	\ContinuedFloat
	\caption{SMF Pseudocode Cont.}
	\begin{algorithmic}[1]
\algrestore{smf}
\Procedure{Calculate\_Share()}{}
\State $selector \gets Epoch \pmod{2}$;
\State $heap \gets \emptyset$;
\State $simulatedCapacity = Capacity$;
\State $simulatedShare \gets 0$;
\State $result \gets 0$;
\For{$i \gets 1, NumberOfUsers$}
    \If{$User.DemandEpoch[selector] = Epoch - 1$}
        \State $\Call{insert}{heap, User.demand[selector]}$;
    \EndIf
\EndFor
\State $simulatedShare \gets \floor*{\frac{simulatedCapacity}{heapSize}}$;
\While{$heap.length > 0 \And simulatedCapacity \geq heap.length$}
    \While{$heap[0] < simulatedShare$}
        \State $simulatedCapacity \gets simulatedCapacity - heap[0]$;
        \State $\Call{deleteMin}{heap}$;
    \EndWhile
\State $simulatedCapacity \gets simulatedCapacity - simulatedShare * heap.length$;
\For{$i = 0, heap.length$}
    \State $heap[i] \gets heap[i] - simulatedShare$;
\EndFor
\State $result \gets result + simulatedShare$;
\State $simulatedShare \gets \floor*{\frac{simulatedCapacity}{heap.length}}$;
\EndWhile
\State \Return $result$;
\EndProcedure
\end{algorithmic}
\end{algorithm}

The \texttt{demand} function takes the user id and demand volume as arguments, and starts with updating the state by a call to \textit{update state} function. In order to select the right portion of the circular buffers, a selector variable is initiated after the state is updated. The \texttt{demand} function writes to $D_0$ in odd epochs, and to $D_1$, in even epochs (line $3$).

Next is to check whether the user has already made a demand in the then present epoch. If so, the function returns, and if not, moves on to record the demand. Lastly, the variable for keeping the epoch at which the user made the last demand is updated to be the then current epoch, and the function returns.

The \texttt{claim} function starts with a call to the \textit{update state} function, and initiates the selector variable. In this claim function also, $D_0$ is used in the even epochs, and $D_1$ in odd ones, like it is in the claim function of QMF.

The function continues with updating the user claim epoch. Finally, it returns after assigning the minimum of the share and the user demand to the user account (line $20$), and discounting that amount from the capacity (line $21$).

The \textit{update state} is an \textit{internal} function, as seen above, called by \texttt{demand} and \texttt{claim} functions. It is mainly responsible for updating the epoch (lines $24-25$); and if the epoch needs to be updated, \textit{capacity} (line $26$), and \textit{share} (line $27$) variables along with it.

\clearpage

The \textit{calculate share} function uses $5$ local variables, thus starts with initiating them. First is the selector variable. Next is the heap, which is used to simulate the demand heaps in the conventional algorithm. In order to keep the global \textit{capacity} variable unaltered, a local variable with the name \textit{simulatedCapacity} is used instead. In order to keep the temporary share in between the iterations, another local variable \textit{simulatedShare} is used, and cumulating shares are collected in the local variable \textit{result}, in order to be returned in the final.

There are two main loops in the algorithm. The first loop (lines $36-40$) is responsible for reading the user demands from the demand vector (storage variable), and if the demand is valid (i.e. recorded in the immediately previous epoch, line $37$) writing to the local heap. This means two storage reads (one for \textit{demand epoch} and one for \textit{demand volume}) and a single memory write, the former of which is relatively costly.

Having prepared the local heap, in line $41$, the simulated share of the first iteration is calculated. Lines $42-53$ show the second main loop of the function. The loop runs until either the heap has been emptied (which means that all the demands are satisfiable with the capacity at hand)  or the capacity is less than the number of demands.

Two additional heaps are nested within this loop. In lines $43-46$ the demands that are fully satisfiable, in other words, the demands that are less than or equal to the simulated capacity of the then present iteration, are deducted from the capacity, and the demand, being fully satisfied, is removed from the heap. The loop breaks if and when it encounters the first demand that is greater than the simulated share, since they can only be offered as much as the simulated share.

Instead of taking each demand and deducing one simulated share from the capacity for each, the remaining number of demands is multiplied with the simulated share, and that total is deducted from the capacity in a single step (lines $54-55$). The second nested loop (lines $548-50$), in turn, iterates over the local  demand heap, and deducts simulated share from the remaining demands. The simulated share is cumulated in the result variable (line $51$), and then recalculated for the next loop (line $52$). When the outer loop termiates, the result variable is returned (line $54$) to the \textit{update state} function.

The functions to insert to and remove from the local heap are what is called \textit{pure} functions in Solidity. They do not read from storage variables, in addition to not writing on them, which renders this category of functions the least costly. The main cost stems from the first outer loop, leading to the limitation on the number of demands, and consequently, on the number of users.

\section{Procedure and Parameters}
\label{procedure}

In the tests we run to measure the performance of our implementations, we chose the bottleneck parameter values to test and keep them identical for all of the restructured algorithms' tests, and set the remaining parameters in relation to them. The parameters in question here are the ones that decides the number of iterations of the \texttt{calculate\_share} functions' loop. In the cases of QMF and WQMF this is the \textit{Quanta} value ($q$), and in SMF and WSMF it is the \textit{number of users} ($n$). We have run our tests with $10$, $50$, $100$, $250$, $500$, and $1000$ quanta values for W/QMF, and with $10$, $50$, $100$, $250$, $500$, and $1000$ users for W/SMF, and observed in each case how the cost scales with these growing values. The chosen values for each parameter can be seen more explicitly in Tables \ref{wqmf_parameters} and \ref{wsmf_parameters} for W/QMF and W/SMF, respectively.

\begin{table}[ht]
    \centering
    \caption{The parameters and their values used in the tests for QMF and WQMF.}
    \begin{tabular}{rrp{3.5cm}}
    Parameter       & Value & Definition \\
    \hline
    \hline
    Quanta          & $Q$       & Maximum demand volume\\
    \hline
    Number of Users & $1000$    & The number of users in the system\\
    \hline
    Epoch Capacity  & $500 \cdot Q$   & The amount to be distributed for each epoch\\
    \hline
    Epoch Span      & $2000$    & The duration of an epoch in number of blocks\\
    \hline
    Demand Interval & $[1,Q]$   & The interval which the demands are uniformly drawn from\\
    \hline
    Weight Interval & $[1,10]$  & The interval which the weights are uniformly drawn from\\
    \hline
    \end{tabular}
    \label{wqmf_parameters}
\end{table}

\begin{table}[ht]
    \centering
    \caption{The parameters and their values used in the tests for SMF and WSMF.}
    \begin{tabular}{rrp{3.5cm}}
    Parameter       & Value & Definition \\
    \hline
    \hline
    Number of Users & $n$       & The number of users in the system\\
    \hline
    Epoch Capacity  & $20n$     & The amount to be distributed for each epoch\\
    \hline
    Epoch Span      & $2n$      & The duration of an epoch in number of blocks\\
    \hline
    Demand Interval & $[15,35)$ & The interval which the demands are uniformly drawn from\\
    \hline
    Weight Interval & $[1,10]$  & The interval which the weights are uniformly drawn from\\
    \hline
    \end{tabular}
    \label{wsmf_parameters}
\end{table}

In W/QMF, the demands are drawn from a discrete uniform distribution in $[1,Q]$ interval, and the capacity is set to $500 \cdot Q$, which is slightly less than the expected average ($\mathbb{E}[d_u] = \frac{1 + Q}{2}, u \in U$) of the uniform distribution in the interval $[1,Q]$. We deliberately introduced this shortage in order for the tests to allow the cases where the total volume of the demands exceed the capacity at hand. No further shortage or abundance of resources is forced into the tests by other parameters. Similarly, in W/SMF, the demands are drawn from the $[15,35)$ discrete interval uniformly, and the capacity is set to $20 \cdot n$, offering each user slightly less than the expected average ($ \mathbb{E}[d_u] = 24.5, u \in U$), in order to allow the tests to include capacity exceeding total demand volume cases.

In the cases of constant weights, the weights are drawn from a uniform distribution in the $[1,10]$ discrete interval ($w \in \mathbb{N}$).

\section{Results}
\label{results_2}

The results for CMF and W/QMF and W/SMF are presented in the following subsections. The data are available at \cite{metin2020faucet}.

\subsection{QMF and SMF Results}
\label{qmf_smf_results}

We summarize the results in three tables, in which we the present the average costs for the \texttt{demand} (Table \ref{demand_avg}), the average costs for the \texttt{claim} (Table \ref{claim_avg}), and the maximum costs for the \texttt{update\_state} (Table \ref{updatestate_max}) functions. The reason for preferring maximum values instead of average in the latter case is that it is more convenient to consider the worst case scenarios rather than the average case, since this is the main bottleneck in all the algorithms. Moreover, the distribution of the cost of this function is widely skewed, the first invocation at each epoch being several orders of magnitude larger than the remaining invocations of the function, rendering the arithmetic average misrepresentative. Finally, for reasons of fairness, this cost is refunded to the user, since being the first to invoke a \texttt{claim} or a \texttt{demand} in a given epoch is hardly a burden that may be fairly loaded on a single random user. Thus, it is not really a cost for the users to shoulder, but rather for the system itself, and the only important concern for this cost is to keep it within the boundaries of the block gas limit.

\renewcommand{\thefootnote}{\fnsymbol{footnote}}


\begin{table}[hhtbp]
\begin{minipage}{\textwidth}
	\caption{Maximum Gas Cost for \textit{Update State}}
	\scalebox{0.98}{
  	\begin{tabular}{r|r|r|r|r|r|r|r}
    \toprule
          q & \multicolumn{1}{c|}{QMF} & \multicolumn{1}{c|}{WQMF} & & n & \multicolumn{1}{c|}{SMF} & \multicolumn{1}{c|}{WSMF-C} & \multicolumn{1}{c}{WSMF-R} \\
    \midrule
    10  & 61,083    & 56,528  & & 10    & 79,859    & 141,159   & 114,289 \\
    \midrule
    50  & 100,353   & 109,094 & & 50    & 237,336   & 693,835   & 724,976 \\
    \midrule
    100 & 163,878   & 128,952 & & 100   & 450,576   & 1,480,234 & 1,438,054 \\
    \midrule
    250 & 317,493   & 404,156 & & 250   & 1,277,618 & 5,247,693 & 4,831,477 \\
    \midrule
    500 & 602,169   & 714,624 & & 500   & 2,611,722 & \footnotemark[2] & \footnotemark[2] \\
    \midrule
    1000& 1,087,829 & 1,285,120 & & 1000& 5,257,236 &     \footnotemark[2]       & \footnotemark[2] \\
    \bottomrule
    \end{tabular}%
	}
	\label{updatestate_max}%
\end{minipage}
\end{table}%


\begin{table}[htbp]
\begin{minipage}{\textwidth}
  \caption{Average Gas Cost for \textit{Demand}}
  	\scalebox{0.87}{
    \begin{tabular}{r|r|r|r|r|r|r|r|r|r}
    \toprule
          q & \multicolumn{1}{c|}{QMF} & \multicolumn{1}{c|}{WQMF} & & n  & \multicolumn{1}{c|}{SMF} & \multicolumn{1}{c|}{WSMF-C} & \multicolumn{1}{c}{WSMF-R} & \multicolumn{1}{c|}{AMF} & \multicolumn{1}{c}{WAMF} \\
    \midrule
    10    & 66,751 & 74,544 & & 10  & 65,101 & 60,161 & 75,837 & 70,245 & 79,732 \\
    \midrule
    50    & 67,754 & 75,878 & & 50  & 65,101 & 60,161 & 75,837 & 67,351 & 77,135\\
    \midrule
    100   & 69,008 & 77,150 & & 100 & 65,101 & 60,161 & 75,837 & 66,989 & 76,835\\
    \midrule
    250   & 72,701 & 79,902 & & 250 & 65,101 & 60,161 & 75,837 & \footnotemark[1] & \footnotemark[1]\\
    \midrule
    500   & 77,743 & 83,066 & & 500 & 65,101 &   \footnotemark[8]    & \footnotemark[8] & 66,700 & 71,365\\
    \midrule
    1000  & 84,817 & 88,440 & & 1000& 65,101 &   \footnotemark[8]    & \footnotemark[8] & \footnotemark[1] & \footnotemark[1] \\
    \bottomrule
    \end{tabular}%
	}
  \label{demand_avg}%
\end{minipage}
\end{table}%

\begin{table}[htbp]
\begin{minipage}{\textwidth}
  \caption{Average Gas Cost for \textit{Claim}}
  \scalebox{0.87}{
    \begin{tabular}{r|r|r|r|r|r|r|r|r|r}
    \toprule
          q	& \multicolumn{1}{c|}{QMF} & \multicolumn{1}{c|}{WQMF} & & n & \multicolumn{1}{c|}{SMF} & \multicolumn{1}{c|}{WSMF-C} & \multicolumn{1}{c}{WSMF-R} & \multicolumn{1}{c|}{AMF} & \multicolumn{1}{c}{WAMF} \\
    \midrule
    10    & 56,122 & 56,721 & & 10 & 56,142 & 56,641 & 57,031 & 46,800 & 46,643\\
    \midrule
    50    & 56,121 & 56,719 & & 50  & 56,142 & 56,639 & 57,533  & 42,240 & 44,852\\
    \midrule
    100   & 56,120 & 56,720 & & 100 & 56,142 & 56,640 & 57,532  & 42,114 & 44,763\\
    \midrule
    250   & 56,120 & 56,719 & & 250 & 56,142 & 56,641 & 57,531 & \footnotemark[1] & \footnotemark[1] \\
    \midrule
    500   & 56,119 & 56,719 & & 500 & 56,142 &    \footnotemark[8]   & \footnotemark[8] & 42,047 & 45,319\\
    \midrule
    1000  & 56,119 & 56,719 & & 1000& 56,142 &    \footnotemark[8]   & \footnotemark[8] & \footnotemark[1] & \footnotemark[1] \footnotetext[0]{$*$\textcolor{white}{$*$}Tests that are not carried out} \footnotetext[0]{$^{\dag \textcolor{white}{\dag}}$ Gas cost exceeds $8,000,000$ block gas limit} \footnotetext[0]{$^{\dag\dag}$ Tests that are not completed because \textit{update\_state} function exceeded the block gas limit}\\
    \bottomrule
    \end{tabular}%
	}
  \label{claim_avg}%
\end{minipage}
\end{table}%

\renewcommand{\thefootnote}{\arabic{footnote}}

As seen in Tables \ref{demand_avg} and \ref{claim_avg}, the cost of demand and claim functions are contained well within the block gas limit, being $2$ orders of magnitude below it, and showing low variability. In fact, the cost of the \texttt{demand} function for W/SMF is constant (i.e. $\sigma = 0$) for both within and between the trials, and between the tests with different numbers of users.

\clearpage

As for the \textit{update state} maximums, Table \ref{updatestate_max} reveals, the growth of the cost tends to linear, and the reported values are well contained within the block gas limit. The missing values in Table \ref{updatestate_max} are due to the fact that, WSMF exceeds the block gas limit for these number of user values, and the presented tests in the previous study \cite{metin2022} were considered sufficient in W/AMF.

\chapter{DISCUSSION}
\label{discussion}

The present dissertation investigates Max-min Fairness distribution scheme in the block- chain ecosystems context over its implementations as blockchain faucets. To point at the generality of the investigation, we should first denote that the algorithms developed hereby are not specific to faucet systems, and they can easily be adopted to any system within the blockchain context that needs to utilise some kind of a distribution scheme, without running into problems specific to blockchain systems. In this sense, the faucet mechanism should be taken as an example and not as the main subject of investigation. The present implemen- tations of Max-min Fairness, for example, can be built within a scheduler operating on a blockchain system.

Nevertheless, the utilities of blockchain faucets are rich. Although they have been con- ceived as free cryptocurrency services for test networks, the function of blockchain faucets should not be taken limited to this use case. For instance, faucets may also serve as distri- bution mechanisms for systems that run on donations (e.g. election rallies), where public transparency, responsibility, incentivisation, and participation are indispensible properties. This kind of a distribution mechanism lends these projects the opportunity to be publicly transparent, and make commitments (e.g. declaring the weights for the expenditure items) prior to raising funds, since the system assures the enforcement of declared commitments, by the virtue of its immutability. Another example may be utilising fair faucets for the distribution of governance tokens in collectively governing communities. The fairness of distribution, in this case, would account for the fairness of decision making processes.

For reasons of simplicity, in the present study the resource to be distributed is repre- sented only over its quantity, with an integer. However, in the contracts we developed, we implemented a simple function to allow users to withdraw from their balances, which can easily be modified to convert the data type to another desired one. For example, a standard token template can be included in the contract and the balance, which is represented as a simple integer, might be converted to the desired token type in the withdraw function. In this sense the contracts presented hereby are compatible with all token standards.

The main bottleneck, and thusly the main performance metric of the present dissertation is the gas consumption, and this is arguably a natural approach for studies on blockchain systems. However, the results presented in this study are not to be taken for their absolute values. Over time, changes in the charges, or low level efficiency improvements in coding or compilation may be introduced, leading to lower transaction costs. The aim of our approach is to demonstrate the availability, and the \textit{cost structure} of the Max-min Fairness algorithm, and its different implementations.

Accordingly, the present dissertation demonstrates, over the failure of CMF to support more than $10$ users, that it is not feasible for Max-min Fairness scheme to be implemented in the blockchain context as it is implemented in the conventional computational settings. In principle, because of the block gas limit, blockchain systems are not well suited for algorithms, which cannot be efficiently distributed to be processed by multiple computing parties, with partial data, and asynchronously. A single transaction to carry out a function with heavy computational burden is not a working strategy while developing software for blockchain systems.

This is in accordance with the distributed nature and the philosophy of the blockchain systems. In contrast with the centralised systems, blockchains aim to distribute both the work and the control among its users. For this reason, they are \textit{incentive driven}, as opposed to centralised systems, which are \textit{authority driven}. That is to say, centralised systems rely on an authorised component (operating system kernels, load balancers, web servers etc.) to carry out the computation; whereas blockchain systems rely on incentivising its users to operate the system in a way that the outcome will turn out to be the desired computation.

It should be noted that the total cost of the \texttt{claim} function in W/AMF is obtained by multiplying the values presented in Table \ref{claim_avg} by the number of calls to the function with the necessary number of calls, which is a function of the distribution of the demands, and may vary among the users with different demands, since in W/AMF total claim process may take more than a single call in each epoch. For example, in the extreme cases where all demands are below or all demands are above the available average (i.e. $\forall u$  $d_u \leq \frac{c}{n}$, or $\forall u$ $d_u \geq \frac{c}{n}$ ) the algorithm takes a single iteration, assigning each user their demands in the former case, and $\frac{c}{n}$ in the latter. In the simulations we run where the demands were uniformly distributed, we observed that the algorithm most usually takes $3$ iterations, and assumed this number as the number of rounds in the W/AMF tests for this reason. A finer analysis on the distribution of the number of iterations Max-min Fairness takes, over the distribution of demands is well beyond the scope of the present study, and to our best efforts, we also were not able to find a study on possible upper-bounds related to this number.

The maximum number of users we reached that WSMF can support under $8.000.000$ gas limit is $250$. Nevertheless the algorithm can be optimised further in the low level in order to decrease the cost (e.g. reduce the size of the variables) and allow for higher numbers of users. We did not undertake such an endeavour for two reasons: First, the main aim and the scope of the present dissertation is to demonstrate the cost structure, rather than to pro- vide tight bounds for the cost. Second, the exact cost of the operation of the \texttt{calculate share} (and consequently the \texttt{update state}) function is again a function of the dis- tribution of the demands. Prospective studies may improve on the absolute values of gas consumption in each algorithm, taking the cost growth structure analyzed and presented here as a basis for their design.

Another lane for future studies would be changing the capacity replenishment policy. In the present dissertation, the capacity is replenished by a constant quantity $C$ at the begin- ning of each epoch. Instead, the tests can be run with varying quantities of replenishment over time, possibly according to some function of epoch number (i.e. $C = f(E)$). In the same vein, the distribution of the user demands may be manipulated in order to observe the outcomes of replenishment and weighting policies.

In the present dissertation we employed two basic weighting policies. In the first case, which may said to be the trivial case, we randomly assigned \textit{constant} weights to users. In the second case, we assigned each user the multiplicative reciprocals of the total sums that they demanded up to and including the then present epoch. The first policy is important for the study because it serves as a base case for comparison for the added complexity of calculating weights with different policies. For example, as explained in Section \ref{wamf}, for the implementation of the second weighting policy, since floating point numbers are not suported in Solidity, and are required to represent multiplicative reciprocals, we developed a custom method for these calculations, and the costs of the first weighting policy served as a reference for comparison for the added cost of this computation in the second policy.

Weighting the users inversely with the total sum of their previous demands implies a policy for incentivising the users to make minimal demands that can satisfy their needs, in order not to be disadvantageous in the long run. Moreover, it can also serve for the long term fairness of the distribution. We were not able to implement the same policy for WQMF, because the weight of the user is needed for the calculations in the demand function. Since the demand interval is limited and the total demand is a monotonous non-decreasing function, and we need the ratio of demand and weight in the registering of the demand, it would lead to the demands piling up to the lower ends of the demand array, rendering the system inoperable.

An alternative policy for WQMF similar to reciprocating the total sum of previous demands could be reciprocating the total sum of a finite number of most recent demands. The added cost of such a policy is affordable, first of all because this cost will be reflected in the demand function, rather than the calculate share function, which is the original bottleneck in WQMF. Depending on the decision of the recency measure, for $n$ recent demands, each user would be allocated a circular buffer of size $n$, and have $n$ additional storage reads and $1$ additional storage write in each call of the demand function. Other than that the algorithm operates identically with the present implementations. Although similar to the second policy we implemented, such a policy would emphasise \textit{recent} rather than the \textit{total} history of the demands of the user, and its implications on the user incetivisation, and ramifications on the user behaviour would be different. Unfortunately, further investigation of the topic is out of the scope of the present dissertation.

The faucet algorithms presented in this study are designed for single resource distri- bution. For the prospective studies we might further propose focusing on multi-resource distribution problems. One way would be keeping each resource type separate and distribute them independently, with the algorithms developed in the present dissertation. For such a policy, no alteration in the present algorithms would be required. The user might deploy multiple contracts and distribute each resource with one. In fact, this would be the right way to proceed if the user intends to use W/QMF or W/SMF, since multiple resource kinds would be sharing the available computational budget, leading for reduced quanta interval support for W/QMF, and reduced number of user support for W/SMF. This is because of the limitation on the available iterations of the loops of these algorithms, since the number of available iterations will be divided among the resource types. For example, under the same block gas limit, doubling the type of resources leads to halving the available quanta interval in W/QMF, or halving the maximum number of supportable users in W/SMF; tripling leads to reducing the sizes to one-thirds.

On the other hand, if the resource types will be considered and distributed in relation to each other, this is a question of policy. In \cite{ghodsi2011dominant}, Ghodsi et al. develop Dominant Resource Fairness, and show that it satisfies fairness demands of multiple resource distribution systems, and it has gained wide acceptance in the literature. The author of the present dissertation is convinced that DRF would be the right method for distributing multiple resources while establishing fairness among them. DRF can be adapted to blockchain systems most efficiently over W/AMF, since the only additional computational burden of calculating share for an additional resource will be two storage reads, a division, and a storage write. This is due to the fact that, in W/AMF, the share for each round is calculated simply by dividing the capacity for a given resource by the total number  or total weight of its demanders, respectively. One storage read is for reading the capacity for the resource, one for reading the total number or the total weight of its demanders, the division for dividing the former by the latter, and storage write to write the result to the \textit{share} or \textit{unit share} variable, again, respectively. This calculation process will be carried out for each individual variable, adding up to the total cost of the \texttt{update state} function.

DRF can also be adapted with W/QMF and W/SMF, but being subject to the limitations described, they would not scale as well as W/AMF could. The total available gas budget would be divided among the resouces, each one being capable of iterating a loop downscaled linearly by the number of resources. In W/QMF this would lead to a trade-off between the available quanta value and the number of resources. Since in DRF, the distribution is done over the percantage quantity of the demand volume to the resource capacity, and Solidity does not support floating point variables implicitly, this translates to the floating point precision the system can support. In W/SMF, on the other hand, the same trade-off is encountered between the number of resources, and the number of users the system can support.

For all of the algorithms, the extra gas cost of ordering the demands for their relative dominance should be handled in the \texttt{demand} function, rather than in \texttt{update state}. This way, the cost will be distributed over the users, preventing a possible bottleneck that would arise for the necessity of iterating over all the demanders in a single loop in the \textit{update state} function. The gas cost analysis of this process is out of the boundaries of the present dissertation. Nevertheless, relative to the block gas limit, the gas cost of \texttt{demand} function is low, at least by two orders of magnitude, and a possible implementation, although another possible bottleneck, should allow for a reasonable number of resources; probably more than the already existing bottlenecks would allow. At this point, we contend with this rough estimation, and leave the numerical analysis to the future study.

An additional caveat for utilising DRF in the blockchain context is the accumulating nature of the distribution process. In the original setting where DRF is developped to address (e.g. resource scheduling in cluster computation), the resources to be distributed are utilised immediately once they are allocated (e.g. CPU or memory). In our distribution process, the resources have not been constrained to this precondition, and they accumulate as long as the user prefers to keep them. This might pose a problem in the DRF context.

More specifically, it may leave the algorithm vulnarable for certain strategies to be developped to short circuit the relative fairness constructs of the algorithm. For example, the users may prefer to maximise their share in only a limited number of resources for a given epoch. In order to achieve this, they may submit inflated demands for the remaining resources, forcing the real targeted resources to sink in the lower priority distribution loops, where their chances of getting larger shares will be higher. Then in the following epochs, other variables will be prefered. This way, by targetting to maximise each variable in different epochs, and accumulating them, the user may get better-off in the middle run.

On the other hand, utilisation of such strategies by all users forces the maxmin algorithm to collapse to the trivial distribution scheme in the higher priority distribution loops, where each of the $n$ users obtain $\frac{1}{n}$ of the resources, since the inflated demands end up degenerating the process. A method for overcoming such problems is to introduce time-to-live (ttl) variables for the distribution tokens, but the discussion of this, also, is not in the scope of the present dissertation, and left for the future studies.

\chapter{CONCLUSION}
\label{conclusion}

In the present dissertation, we addressed the problem of fair distribution of shared resources within the blockchain systems context. We worked on the intrinsic resources of blockchains, and developed faucets as smart contracts, running different implementations of Max-min Fairness Algorithm, which is traditionally accepted in the literature for realising fairness.

It has been demonstrated that the Max-min Fairness algorithm, as it is implemented in the conventional programming contexts, cannot support a public system because of the scaling of its gas cost structure. Autonomous and restructured implementations of the algorithm are offered as solutions, and the tests have shown that these implementations can support wide public use of the system without running into block gas limit exhaustion problem. As the results reveal, the \texttt{demand} and \texttt{claim} functions exhibit low variability among the algorithms, and all are efficient with respect to the block gas limit, being several orders of magnitude below it.

The algorithms presented hereby bear relative advantages to each other and each one is optimal for a different use case. In the cases where multiple time-critical calls on the client side is acceptable, W/AMF works with the lightest computational burden. On the other hand, if the restrictions on demand volume and weights do not conflict with the operational requirements, W/QMF stands out to be costwise the most efficient as compared to their counterparts. Finally, in cases where the necessary support for the number of users are within the limits presented here, W/SMF present the richest functionality in the most efficient way.

We presented expirical results to support our hypotheses and compared our algorithms numerically in their related sections and subsections. After demonstrating our claims empir- ically, we introduced a comprehensive conceptual discussion in the penultimate chapter, and we presented our ideas and projections on follow-up studies, pointing out to the challenges they bear, and possible outcomes they may lead.

To conclude, blockchain systems are gaining ever increasing emphasis in the modern
day technology. It is reasonable to expect their wide use in daily life in a not much distant
future. The present dissertation is intended to be a first step to constructing a means to
develop and analyse blockchain server utilities, in isolation from the proofing mechanism. It
focuses on the strcuture of the scaling cost and the incentive involved in the system over an
examplary utility. It is, therefore, presented both for its experimental contributions and as a
working model for studying blockchain system utilities in general.

\bibliographystyle{styles/fbe_tez_v11}
\bibliography{references}

\appendix

\chapter{AMF Pseudocode}
\label{amf_pseudo}

\begin{algorithmic}[1]
\Procedure{Update State}{$Offset, BlockNumber, Epoch, EpochSpan, RoundSpan$}
\State $selector \gets Epoch ~~mod~~(2)$;
\If{$Epoch < \floor*{ \frac{BlockNumber - Offset}{EpochSpan} }$}
	\State $Epoch \gets \floor*{ \frac{BlockNumber - Offset}{EpochSpan} }$;
	\State $Round \gets \floor*{ \frac{(BlockNumber - Offset) ~~mod~~ (EpochSpan)}{RoundSpan} }$;
	\State $Capacity \gets Capacity + EpochCapacity$\;
	\State $Share \gets \floor*{Capacity / TotalWeight[selector]}$;
	\State return;
\EndIf
\If{$Round < \floor*{ \frac{(BlockNumber - Offset) \% ES}{RS} }$}
	\State $Round \gets \floor*{ \frac{(BlockNumber - Offset) ~~mod~~ (EpochSpan)}{RoundSpan} }$;
	\State $Share \gets Capacity / TotalWeight[selector]$;
	\State return;
\EndIf
\State return;
\EndProcedure
\Procedure{Demand}{$User, Volume$}
\State \Call{updateState}{$Offset, BlockNumber, Epoch, EpochSpan, RoundSpan$}
\State $selector \gets (E + 1)~~mod~~(2)$;
\If{$User.demandEpoch[selector] \neq Epoch$}
	\State $User.demand[selector] \gets Volume$;
	\State $User.demandEpoch[selector] \gets Epoch$;
	\If{$ResetEpoch < Epoch$}
		\State $TotalWeight[selector] \gets User.weight$;
		\State $ResetEpoch \gets Epoch$;
	\Else
		\State $TotalWeight[selector] \gets TotalWeight[selector] + User.weight $;
	\EndIf
\EndIf
\State return;
\EndProcedure
\Procedure{Claim}{$User$}
\State \Call{updateState}{$Offset, BlockNumber, Epoch, EpochSpan, RoundSpan$}
\State $selector \gets Epoch ~~mod~~(2)$;
\If{$User.demandEpoch[selector] \neq Epoch - 1$ \textbf{or} $Capacity = 0$
					\item[] \hfill \textbf{or} $User.demand[selector] = 0$}
	\State return;
\EndIf
\If{$User.claimEpoch = Epoch$}
	\If{$User.claimRound = Round$}
		\State return;
	\EndIf
\Else
	\State $User.claimEpoch \gets Epoch$;
\EndIf
\State $User.claimRound \gets Round$;
\State $User.balance \gets User.balance + \min{(User.demand[selector], Share * User.weight)}$;
\State $User.demand[selector] \gets User.demand[selector] - \min{(User.demand[selector], Share * User.weight)}$;
\State $Capacity \gets Capacity - \min{(User.demand[selector], Share * User.weight)}$;
\If{$User.demand[selector] = 0$}
	\State $TotalWeight[selector] \gets TotalWeight[selector] - User.weight$;
\EndIf
\State return;
\EndProcedure
\end{algorithmic}

\chapter{WQMF Pseudocode}
\label{wqmf_pseudo}

\begin{algorithmic}[1]
\Procedure{Demand}{User, Volume}\Comment{Make a Demand}
\State \Call{Update\_State()}{};
\State $selector \gets (epoch + 1) \pmod{2}$;
\If{$User.demandEpoch[selector] = Epoch$}
    \State return;
\EndIf
\State $User.demandEpoch[selector] \gets Epoch$;
\State $index \gets \ceil*{\frac{Volume}{User.weight}}$;
\If{$ResetEpoch[selector][index] \neq Epoch$}
    \State $ResetEpoch[selector][index] \gets Epoch$;
    \State $Demands[selector][index] \gets Volume$;
    \State $Weights[selector][index] \gets User.Weight$;
\Else
    \State $Demands[selector][index] \gets Demands[selector][index] + Volume$; 
    \State $Weights[selector][index] \gets Weights[selector][index] + User.weight$;
\EndIf
\State $User.demand[selector] \gets Volume$;
\State $TotalDemands \gets TotalDemands + Volume$;
\State $TotalWeights \gets TotalWeights + User.weight$;
\EndProcedure
\Procedure{Claim}{User}\Comment{Claim User Share}
\State \Call{Update\_State()}{};
\State $selector \gets Epoch \pmod{2}$;
\If{$User.demandEpoch[selector] = Epoch$}
    \State return;
\EndIf
\If{$User.claimEpoch = Epoch$}
    \State return;
\EndIf
\State $User.claimEpoch \gets Epoch$;
\State $share \gets User.weight * UnitShare$;
\State $User.balance \gets min(share, User.demand[selector])$;
\State $Capacity \gets Capacity - min(share, User.demand[selector])$;
\EndProcedure
\Procedure{Calculate\_Unit\_Share()}{}
\State $selector \gets Epoch \pmod{2}$;
\State $cumulativeDemands \gets 0$;
\State $cumulativeWeights \gets 0$;
\For{$i \gets 1, Quanta$}
    \If{$ResetEpoch[selector][i] = Epoch - 1$}
        \State $cumulativeDemands \gets cumulativeDemands + Demands[selector][i]$;
        \State $cumulativeWeights \gets cumulativeWeights + Weights[selector][i]$;
    \EndIf
    \If{$Capacity < cumulativeDemands + i * (totalWeights - cumulativeWeights) $}
        \State \Return $i - 1$;
    \EndIf
\EndFor
\State \Return $Quanta$;
\EndProcedure
\end{algorithmic}

\chapter{WSMF Pseudocode}
\label{wsmf_pseudo}

\begin{algorithmic}[1]
\Procedure{Demand}{User, Volume}\Comment{Make a Demand}
\State \Call{Update\_State()}{};
\State $selector \gets (epoch + 1) \pmod{2}$;
\If{$User.demandEpoch[selector] = Epoch$}
    \State return;
\EndIf
\State $User.demand[selector] \gets Volume$;
\State $User.demandEpoch[selector] \gets Epoch$;
\State $User.totalDemand \gets User.totalDemand + Volume$;
\EndProcedure
\Procedure{Claim}{User}\Comment{Claim User Share}
\State \Call{Update\_State()}{};
\State $selector \gets Epoch \pmod{2}$;
\If{$User.demandEpoch[selector] = Epoch$}
    \State return;
\EndIf
\If{$User.claimEpoch = Epoch$}
    \State return;
\EndIf
\State $User.claimEpoch \gets Epoch$;
\If{$User.demandEpoch[1 - selector] = Epoch$}
    \State $Share \gets UnitShare * \floor*{\frac{Precision}{User.totalDemand - User.demand[selector]}}$;
\Else
    \State $share \gets UnitShare * \floor*{\frac{Precision}{User.totalDemand}}$;
\EndIf
\State $User.balance \gets min(share, User.demand[selector])$;
\State $Capacity \gets Capacity - min(share, User.demand[selector])$;
\EndProcedure
\item[]
\Procedure{Calculate\_Unit\_Share()}{}
\State $selector \gets Epoch \pmod{2}$;
\State $heap[0] \gets \emptyset$;\Comment{Initiate Empty Heaps}
\State $heap[1] \gets \emptyset$;
\State $simulatedCapacity = Capacity * Precision$;
\State $simulatedShare \gets 0$;
\State $simulatedUnitShare \gets 0$;
\State $totalWeight \gets 0$;
\State $result \gets 0$;
\For{$i \gets 1, NumberOfUsers$}
    \If{$User[i].demandEpoch[selector] = Epoch - 1$}
    \State $userWeight \gets \floor*{\frac{Precision}{User.totalDemand}}$
    \State $node \gets \{User.demand[selector], userWeight\}$
    \State $\Call{insert}{heap[0], node}$;
    \State $totalWeight \gets totalWeight + userWeight$;
    \EndIf
\EndFor
\While{$heap[selector].length > 0 \And simulatedCapacity \geq totalWeight$}
    \State $simulatedUnitShare \gets \floor*{\frac{simulatedCapacity}{totalWeight}}$;
    \State $result \gets result + simulatedShare$;
    \While{$heap[selector].length > 0 $}
        \State $simulatedShare \gets heap[selector][0].weight * simulatedUnitShare$;
    \If{$simulatedShare = 0$}
        \State $totalWeight \gets totalWeight - heap[selector][0].weight$;
        \State $\Call{deleteMin}{heap[selector]}$
        \ElsIf{$heap[selector][0].volume \leq simulatedShare$}
            \State $simulatedCapacity \gets simulatedCapacity - heap[selector][0].volume$;
            \State $\Call{deleteMin}{heap[selector]}$;
        \Else 
        \State $node \gets \{heap[selector][0].volume - simulatedShare,$ \hfill
            \item[] \hfill $heap[selector][0].weight\}$;
            \State $\Call{deleteMin}{heap[selector]}$;
            \State $\Call{insert}{heap[1 - selector], node}$;
    \EndIf
    \State $selector \gets 1 - selector$;
    \EndWhile
\EndWhile
\State \Return $result$;
\EndProcedure
\end{algorithmic}

\end{document}